\documentclass[fleqn,usenatbib]{mnras}

\usepackage{newtxtext,newtxmath}

\usepackage{xcolor}
\usepackage{ulem}
\usepackage[T1]{fontenc}

\DeclareRobustCommand{\VAN}[3]{#2}
\let\VANthebibliography\thebibliography
\def\thebibliography{\DeclareRobustCommand{\VAN}[3]{##3}\VANthebibliography}

\usepackage{graphicx}	
\usepackage{amsmath}	
\usepackage{amssymb}	
\usepackage{physics}

\title[Local measurement of growth rate from 6dFGS]{A local measurement of the growth rate from peculiar velocities and galaxy clustering correlations in the 6dF Galaxy Survey}

\author[R. J. Turner et al.]{
Ryan J. Turner,$^{1}$\thanks{E-mail: rjturner@swin.edu.au}
Chris Blake,$^{1}$
Rossana Ruggeri$^{2,1}$
\\
$^{1}$Centre for Astrophysics and Supercomputing, Swinburne
University of Technology, Hawthorn, VIC 3122, Australia\\
$^{2}$School of Mathematics and Physics, The University of Queensland, Brisbane, QLD 4072, Australia.\\
}

\date{Accepted XXX. Received YYY; in original form ZZZ}

\pubyear{2022}

\begin{document}
\label{firstpage}
\pagerange{\pageref{firstpage}--\pageref{lastpage}}
\maketitle

\begin{abstract}

Galaxy peculiar velocities provide an integral source of cosmological information that can be harnessed to measure the growth rate of large scale structure and constrain possible extensions to General Relativity. In this work, we present a method for extracting the information contained within galaxy peculiar velocities through an ensemble of direct peculiar velocity and galaxy clustering correlation statistics, including the effects of redshift space distortions, using data from the 6-degree Field Galaxy Survey. Our method compares the auto- and cross-correlation function multipoles of these observables, with respect to the local line of sight, with the predictions of cosmological models. We find that the uncertainty in our measurement is improved when combining these two sources of information in comparison to fitting to either peculiar velocity or clustering information separately. When combining velocity and density statistics in the range $27 < s < 123 \, h^{-1}$ Mpc we obtain a value for the local growth rate of $f\sigma_8 = 0.358 \pm 0.075$ and for the linear redshift distortion parameter $\beta = 0.298 \pm 0.065$, recovering both with $20.9$ per cent and $21.8$ per cent accuracy respectively. We conclude this work by comparing our measurement with other recent local measurements of the growth rate, spanning different datasets and methodologies. We find that our results are in broad agreement with those in the literature and are fully consistent with $\Lambda$CDM cosmology. Our methods can be readily scaled to analyse upcoming large galaxy surveys and achieve accurate tests of the cosmological model.
\end{abstract}




\section{Introduction}

The Hubble-Lemaitre law tells us that galaxies recede from us and that the velocity of this recession is directly correlated to comoving distance. The expansion of the universe drives this behaviour, and we term the collective movement of galaxies away from us due to this expansion as the `Hubble flow'.  Galaxies are not only affected by expansion, however. Fluctuations in the local density field and the gravitational influence of large-scale structure cause galaxies to deviate from the Hubble flow. These deviations are called peculiar velocities.

Peculiar velocities can be described as a Doppler shift in the cosmological redshift of a galaxy. The observed redshift of a galaxy is then a combined estimate of cosmological expansion and peculiar velocity, expressed numerically as
\begin{equation}
    (1 + z_{\rm obs}) = (1 + z_{\rm cos})(1 + \frac{v_{\rm pec}}{c})\,,
    \label{eq:doppler_zeq}
\end{equation}
where $z_{\rm cos}$ is the cosmological redshift of the galaxy and $v_{\rm pec}$ is the peculiar velocity of the galaxy along the line of sight.

Two fundamental ways of testing the standard cosmological model are to study the cosmic expansion history of the Universe over a wide redshift range and to measure the rate of growth of cosmic structure. The growth rate $f = d \ln D / d \ln a$ is the derivative of the logarithm of the linear growth factor $D$ -- the amplitude of the growing mode of the density perturbation equation -- over the derivative of the logarithm of the scale factor $a$. The growth rate can be measured from several probes: galaxy clustering, weak gravitational lensing, and peculiar velocity measurements, among others \citep[e.g.][]{Peebles1980, Percival2009, Weinberg2013, Huterer2015, Ferreira2019}. Cosmic expansion is measured with the use of standardised, or standardisable, references. Type Ia supernovae were used in this way to illuminate the fact that the expansion rate of the universe is accelerating \citep{Riess1998, Perlmutter1999}. In order to fold this cosmic acceleration into a concordance cosmology that evolves according to the general theory of relativity, the cosmological constant $\Lambda$ \citep{Carroll2001} is invoked in the non-baryonic Cold Dark Matter (CDM) model to give rise to our standard model of cosmology, $\Lambda$CDM. While $\Lambda$CDM provides the best fit to our observations of how the Universe has evolved in late times, the inclusion of dark energy (a negative pressure component with equation of state parameter $\omega = -1$) remains phenomenological. The true nature of dark energy, whether in the form of $\Lambda$ or as some modification to gravitational theory on large scales, remains one of the most fundamental questions in cosmology.

Measurements of the cosmic expansion history of the Universe have reached the order of $\sim1$ per cent accuracy \citep{ebosscollab2020}, while measurements of growth are typically an order of magnitude less precise. By comparing results to predictions made by $\Lambda$CDM the model can be stress-tested, and modified theories of General Relativity (GR) or alternatives to dark energy can be explored. Modified GR or a $\Lambda$-induced acceleration cannot be discerned from one another by probes of the cosmic expansion alone, but can cause different behaviours in the growth rate for fixed expansion -- a stronger gravitational force will mean a larger growth rate, and certain models of modified GR predict a growth rate that is scale-dependent. Measurements of the growth rate of structure can thus be used to differentiate between $\Lambda$, alternatives to dark energy, and modifications to GR that would otherwise be degenerate with one another. 

Additionally, the growth rate can be written in terms of the matter density parameter $\Omega_m$ and the growth index $\gamma$,
\begin{equation}
    f(z) = \Omega_m(z)^\gamma\,.
\end{equation}
The growth index value is set by the chosen theory of gravity. In the $\Lambda$CDM model of cosmology, $\gamma = 6/11$ \citep{Silveira1994, Wang1998, Linder2005}. In the Dvali, Gabadadze \& Porrati braneworld model \citep[DGP;][]{Dvali2000}, $\gamma = 11/16$  \citep{Linder2007, Gong2008}, and in $f(R)$ gravity models the growth rate $f$ is shown to exhibit scale dependence \citep{Hu2007}. If we are able to more accurately measure the growth rate of cosmic structure, then we can restrict any allowed deviations from the standard cosmological model and place tight constraints on proposed alternative theories of gravity.

In our study, we self-consistently combine two observable probes of Equation \ref{eq:doppler_zeq}.  First, the peculiar velocities of galaxies which may be measured directly if the total redshift of a galaxy can be combined with a redshift-independent measure of distance.  Second, peculiar velocities cause the observed density field of galaxies to be distorted along the line of sight in redshift space \citep{Kaiser1987}. The auto- and cross-correlations between the galaxy velocities and redshift-space positions are sensitive to the combined parameter $f\sigma_8$, the normalised growth rate, where $\sigma_8$ describes the amplitude of the matter power spectrum on the scale of $8$ h$^{-1}$ Mpc. 

These redshift space distortions (RSD) are the physical manifestation of the effect described by Equation \ref{eq:doppler_zeq}, which induces an anisotropy in the redshift-space clustering signal. This can be described by the linear redshift distortion parameter, $\beta = f/b$, where $f$ is the growth rate and $b$ is the linear galaxy bias which describes how galaxies trace the underlying matter density field.
Cosmological information can be extracted from RSD by performing a multipole expansion of the galaxy-galaxy auto-correlation function and the galaxy-velocity cross-correlation function, with respect to the local line of sight.

The velocity field is more sensitive to fluctuations on larger scales than the density field. This can be directly seen in the relationship between the velocity modes $\tilde{v}(k)$ and density modes $\tilde{\delta}(k)$ in Fourier space, $\tilde{v}(k) \propto \tilde{\delta}(k)/k$, where the additional $1/k$ factor upweights velocities at larger scales with respect to the density contrasts. The peculiar velocity power spectrum thus traces density modes on large scales, while RSD power is dominated by contributions on smaller scales \citep{Said2020}. These approaches can be combined to produce more accurate results than from either method alone \citep[eg.][]{Koda2014, Howlett2017}. Peculiar velocities are a direct probe of mass fluctuations on the largest scales and encode information about the growth of cosmic structure and gravitational physics, parameterised by $f\sigma_8$.

The normalised growth rate can be measured from a joint analysis of the local density and velocity fields \citep{Strauss1995} where the peculiar velocity field is reconstructed from the density field \citep{Nusser1994, Hudson1995, Pike2005, Davis2011, Carrick2015, Boruah2020, Said2020}, from RSD \citep{Beutler2012, Alam2016}, or from the velocity power spectrum either directly \citep{Howlett2MTF}, via the momentum power spectrum \citep{Howlett2019, Qin2019} or using a maximum-likelihood method \citep{Johnson2014, Huterer2017, Adams2017, Adams2020}. \citet{Adams2020} include the galaxy-velocity cross-power spectrum in their analysis, making this the most comprehensive of this particular form of analysis.

In \citet{Turner2021} we introduced a method of measuring the local growth rate using an optimal combination of peculiar velocity statistics and galaxy clustering statistics, obtained through the auto- and cross-correlation functions of the peculiar velocity field and the galaxy overdensity field, which we tested using N-body simulations.  In this study, we aim to improve this analysis by harnessing the cosmological information from RSD through a multipole expansion of the galaxy-galaxy power spectrum and the galaxy-velocity cross-power spectrum. By doing so we can produce a comprehensive analysis framework for a joint correlation-function study of the galaxy peculiar velocity and density distributions, using data from the \textit{6-degree Field Galaxy Survey} \citep[6dFGS;][]{Jones2004, Jones2005, Jones2009}.

Samples of peculiar velocities are growing rapidly, and now include \textit{Cosmicflows-IV} \citep{Kourkchi2020, Tully2022}, containing over $55,000$ distances, and the \textit{Sloan Digital Sky Survey Peculiar Velocity Catalogue} \citep{Howlett2022}, containing $\sim34{,}000$ distances. Forthcoming surveys conducted on the \textit{Dark Energy Spectroscopic Instrument} \citep[DESI;][]{DESICollab2016} and the \textit{4-metre Multi-Object Spectroscopic Telescope} \citep[4MOST;][]{dejong2019} will have the capability to cumulatively provide peculiar velocity measurements for over $1{,}000{,}000$ galaxies. With such an abundance of velocity data the growth rate $f$ could be measured with approximately $2$ per cent accuracy \citep{Koda2014, Howlett2017}, comparable with measurements of expansion.

Our paper is structured as follows: in Section \ref{sec:data} we discuss the 6dFGS survey data and mock data used in this analysis. In Section \ref{sec:models} we incorporate redshift space distortions into our ensemble of models via the multipoles of the correlation functions. In Section \ref{sec:estimators} we rebuild the correlation function estimators for use with the 6dFGS data and describe how we use these estimators in a chi-squared minimisation procedure. In Section \ref{sec:results} we present the results of this analysis when applied to the 6dFGS mocks and 6dFGS data, fitting to three combinations of statistics, and contrast these results with one another. In Section \ref{sec:6dfcomp} we discuss our results in comparison to other measurements of the local growth rate. In Section \ref{sec:conclusion} we summarise the results of this work and discuss the potential impact that combining future datasets with this method may have for measurements of the growth rate.

\section{6dFGS survey data and mocks}
\label{sec:data}
\subsection{Data}
The 6-degree Field Galaxy Survey is a dual redshift and peculiar velocity survey conducted on the UK Schmidt Telescope using the six-degree field multi-object spectrograph, carried out from 2001 to 2006. The survey covers approximately $17\,000$ deg$^2$ of the southern sky with galactic latitude $|b| > 10^{\circ}$ out to a redshift of $z \approx 0.23$. In this analysis, we work with both 6dFGS samples; the redshift sample, 6dFGSz, and the peculiar velocity sample, 6dFGSv. 

The third and final 6dFGSz sample contains $125\,071$ redshifts, with a median redshift of $z = 0.053$. We use the same galaxy redshift sample in our analysis as \citet{Adams2020}, who applied a redshift cut of $z \leq 0.1$ to the redshift catalogue produced by \citet{Beutler2011} for the 6dFGS baryon acoustic oscillation (BAO) analysis. The BAO catalogue was the result of crossmatching the full 6dFGSz catalogue with galaxies from the Two Micron All Sky Survey Extended Source Catalogue \citep{Jarrett2000} with magnitude $K \leq 12.9$. Galaxies were also excluded from this subsample if they were found in areas of the sky that were lower than 60 per cent complete in 6dFGS. The sample used by \citet{Beutler2011} contained $75\,117$ galaxies after applying selection cuts, the additional redshift cut applied by \citet{Adams2020} reduced this to $70\,467$ galaxies.

The 6dFGSv sample initially contained $11\,287$ galaxies with derived Fundamental Plane (FP) parameters \citep{Magoulas2012}. \citet{Magoulas2012} then excluded galaxies from this sample with signal-to-noise ratios less than $5\Angstrom^{-1}$, velocity dispersions $\sigma_0 \geq 112$ km s$^{-1}$, observed redshifts $cz \geq 16\,120$ km s$^{-1}$ and $cz \leq 3\,000$ km s$^{-1}$ in the CMB frame, as well as excluding galaxies on the basis of visual inspection. This produces a sample of $9\,794$ FP measurements with redshifts $z \leq 0.057$ and with distance errors of approximately $26$ per cent. \citet{Springob2014} further removed several hundred galaxies from this sample that possessed recessional velocities greater than $16 120$ km s$^{-1}$ despite otherwise satisfying the heliocentric redshift limit $z_{\rm helio} = 0.055$. This final sample contains $8\,885$ galaxies.

As shown by \citet{Springob2014}, the error distribution of galaxy offsets to the Fundamental Plane is approximately Gaussian in nature. It follows, then, that the posterior distribution of peculiar velocity measurements inferred from the FP method will be lognormal. This makes the direct use of FP peculiar velocity data in our chi-squared analysis inappropriate. \citet{Springob2014} show that we can instead frame these data in terms of logarithmic distance ratios, $\eta \equiv \log_{10}D(z_{\rm obs}) - \log_{10}D(z_{\rm H})$, where $D(z_{\rm obs})$ is the comoving distance calculated from the observed redshift and $D(z_{\rm H})$ is the comoving distance estimated using the Hubble redshift inferred from the FP method. Peculiar velocity errors that scale with distance translate into a constant error in $\eta$. If peculiar velocity error $\sigma_{pv} = fcz$, where $f$ is some fractional error, then $\sigma_{\eta} = f /\,\ln10$. As $\eta$ increases steeply with decreasing redshift ($\propto 1/z$), low-redshift objects are effectively upweighted.

The logarithmic distance ratio can then be expressed in terms of the peculiar velocity $v_p$ and a redshift-dependent normalisation factor $\alpha$ as, 
\begin{equation}
    \label{logd_to_vp}
    \eta = \alpha v_p \,
\end{equation}
where
\begin{equation}
\label{eqalpha}
    \alpha(z_{\rm obs}) = \frac{1}{\ln(10)}\frac{1 + z_{\rm obs}}{D(z_{\rm obs})H(z_{\rm obs})}\,,
\end{equation}
following \citet{Johnson2014} and \citet{Adams2017}. We neglect a noise term in Equation \ref{logd_to_vp} that arises from the FP relation in the 6dFGSv sample \citep{Magoulas2012}.

\subsection{Simulations}
To test our analysis pipeline and determine errors, we also utilised a set of 600 mock catalogues in our study, originally produced as part of the 6dFGS baryon acoustic peak reconstruction analysis \citep{Carter2018}. These simulations evolved $(1728)^3$ particles in a $1200 \, h^{-1}$ Mpc box using fast Comoving Lagrangian Acceleration (COLA) N-body techniques.  The fiducial cosmological model used for the initial power spectrum of the simulations was $\Omega_m = 0.3$, baryon density $\Omega_b = 0.0478$, Hubble parameter $h = 0.68$, clustering amplitude $\sigma_8 = 0.82$ and spectral index $n_s = 0.96$. The simulation data was output at a snapshot at $z=0.1$, matching the effective redshift of the 6dFGS BAO measurement \citep{Beutler2011}.

The mock galaxy catalogues were generated by populating the dark matter halos of these simulations with a Halo Occupation Distribution (HOD) of central and satellite galaxies calibrated by the observed 6dFGS clustering and number density versus redshift; we refer the reader to \citet{Carter2018} for more details.  Each mock was sub-sampled using the 6dFGS angular selection function.  We note a caveat that the simulation mass resolution was not sufficient to replicate all 6dFGS galaxies at the lowest redshifts.

From each mock redshift catalogue, we generated a mock 6dFGS peculiar velocity catalogue for $z < 0.057$ by selecting the $8{,}885$ most massive central galaxies, and we applied an error $\sigma_\eta = 0.12$ to each PV object (which corresponds to a $26$ per cent error in distance, matching the average properties of 6dFGSv). We generated corresponding random catalogues for the galaxy and velocity samples that could be used in our correlation function estimators.

\section{Modelling}
\label{sec:models}
In this section, we present joint models for the galaxy and velocity correlations in redshift space.

\subsection{RSD models}
\label{ssec:rsdmod}

\begin{figure*}
    \centering
    \includegraphics[width = 0.75\textwidth]{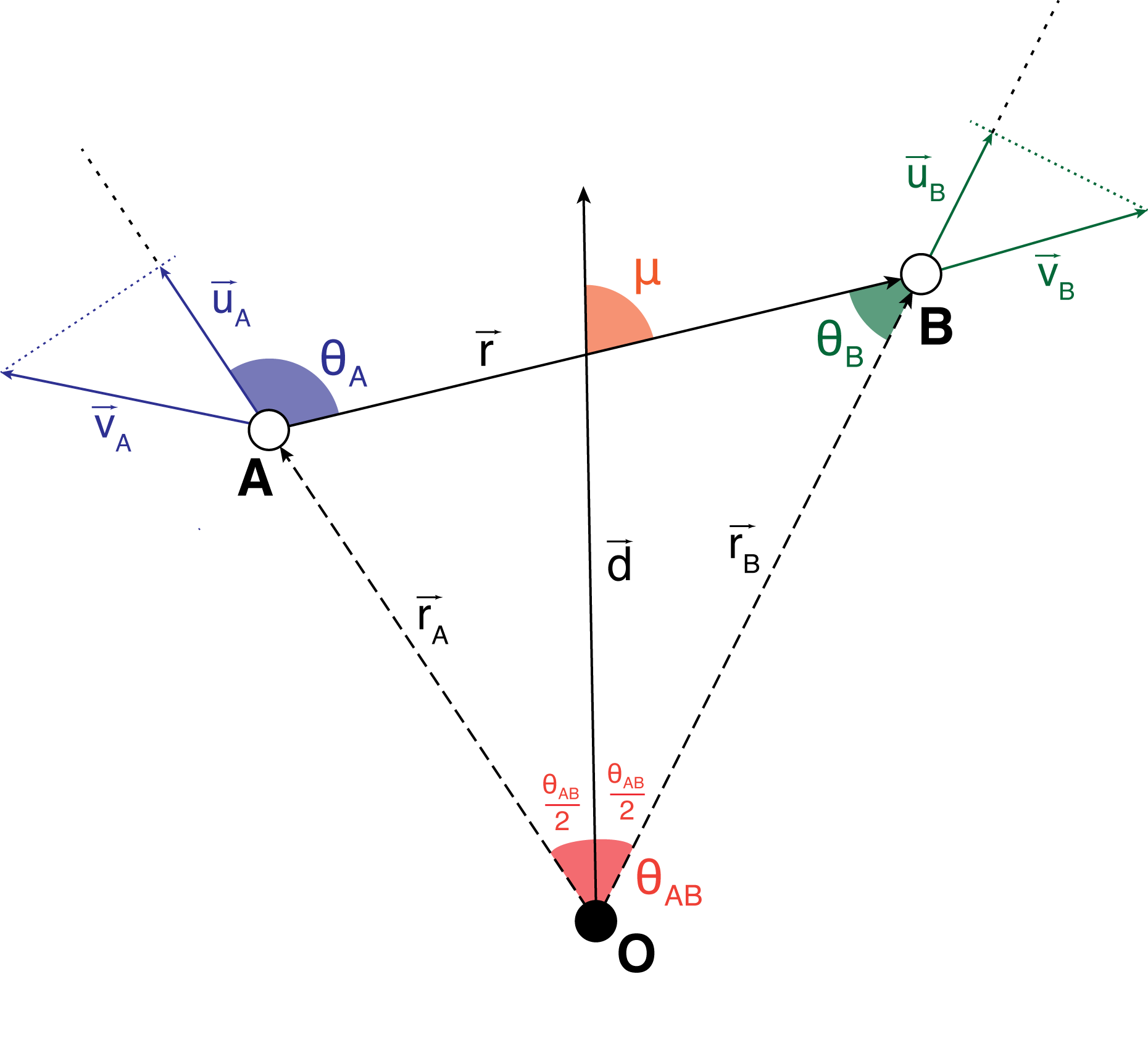}
    \caption{A scenario depicting a pair of galaxies $A$ and $B$ as seen by an observer $O$, and the associated geometry. $\vec{r}_{\rm A}$ and $\vec{r}_{\rm B}$ are position vectors of galaxies $A$ and $B$, respectively, and $\vec{r}$ is the separation vector between the two galaxies. The peculiar velocities of galaxies $A$ and $B$ are represented by $\vec{v}_{\rm A}$ and $\vec{v}_{\rm B}$, and the radial component of these velocities are represented by $\vec{u}_{\rm A}$ and $\vec{u}_{\rm B}$. The angle $\cos\theta_{\rm A}$ = $\hat{\vec{r}}_{\rm A}\cdot\hat{\vec{r}}$, $\cos\theta_{\rm B}$ = $\hat{\vec{r}}_{\rm B}\cdot\hat{\vec{r}}$, and $\cos\theta_{\rm AB}$ = $\hat{\vec{r}}_{\rm A}\cdot\hat{\vec{r}}_{\rm B}$. The angle $\mu$ between the line of sight $\vec{d}$ and the separation vector $\vec{r}$ is also shown, where $\vec{d}$ is chosen so that the angles $\hat{r}_{\rm A}\cdot\hat{d}$ and $\hat{r}_{\rm B}\cdot\hat{d}$ are equal.}
    \label{fig:geom}
\end{figure*}

The standard expression for the Fourier amplitudes of the density field $\delta^s_g$ in redshift space, in terms of the density field $\delta_g$ and peculiar velocity divergence field $\theta = \vec{\nabla}.\vec{v}/aH$ in real space, where $a$ is the cosmic scale factor and $H$ is the Hubble parameter, is
\begin{equation}
    \label{storspace}
    \delta^s_g(\vec{k}) = \delta_g(\vec{k}) - \mu^2\theta(\vec{k})\,,
\end{equation}
where $\mu = \hat{d}\cdot\hat{r}$ is the cosine of the angle between the line of sight $\vec{d}$ and the separation vector $\vec{r}$ between two positions $\vec{s}_1$ and $\vec{s}_2$ \citep{Kaiser1987}. We choose $\vec{d}$ such that $\hat{d}\cdot\hat{s}_1 = \hat{d}\cdot\hat{s}_2$. In Figure \ref{fig:geom} we show a not-to-scale representation of a scenario of an observer $O$ and two galaxies $A$ and $B$ detailing all associated vectors, velocities and angles, including $\mu = \hat{d}\cdot\hat{r}$.

Assuming that the velocity field is derived from linear perturbation theory and that galaxies are linearly biased tracers of mass, $\delta_g(\vec{k})$ and $\theta(\vec{k})$ can be rewritten as $\delta_g(\vec{k}) = b\delta_m(\vec{k})$ and $\theta(\vec{k}) = -f\delta_m(\vec{k})$, where $b$ is the linear galaxy bias factor and $f$ is the growth rate. Using these we can derive linear-theory expressions for the auto- and cross-power spectra of galaxies and velocities as a function of the angle to the line of sight:

\begin{equation}
    \label{ggpower}
    P_{gg}(k , \mu) = \langle \delta^s_g(\vec{k})\delta^{s*}_g(\vec{k})\rangle =  (b + f\mu^2)^2P_{mm}(k)\,,
\end{equation}
\begin{equation}
    \label{gthpower}
    P_{g\theta}(k , \mu) = \langle \delta^s_g(\vec{k})\theta^*(\vec{k})\rangle = (b + f\mu^2)f\,P_{mm}(k)\,,
\end{equation}
\begin{equation}
    \label{ththpower}
    P_{\theta\theta}(k , \mu) = \langle \theta(\vec{k})\theta^*(\vec{k})\rangle = f^2\,P_{mm}(k)\,,
\end{equation}
where $P_{mm}(k) = \langle |\delta_m(\vec{k})|^2 \rangle$ is the matter power spectrum. We generate our matter power spectrum data using the Code for Anisotropies in the Microwave Background \citep[CAMB;][]{Lewis2000,Lewis2011} and set cosmological parameters to the fiducial cosmology of the 6dFGS mock catalogues; $\Omega_m = 0.3$, $\Omega_b = 0.0478$, $h = 0.68$, $\sigma_8 = 0.82$, $n_s = 0.96$. We also use {\sc halofit} \citep{Smith2003} to account for non-linearities. For our analysis of the 6dFGS mocks we compute the matter power spectrum at a redshift of $z = 0.10$, matching the simulation output, and for the analysis of the 6dFGS datasets we compute the matter power spectrum at a redshift of $z = 0.00$.

Equations \ref{ggpower} to \ref{ththpower} can be used to find the non-zero multipoles of the power spectra, $P^\ell(k) = (2\ell + 1)\int^1_0d\mu\,P(k , \mu)L_{\ell}(\mu)$, where $L_{\ell}$ are the Legendre polynomials:
\begin{equation}
    \label{gg0}
    P^0_{gg}(k) = \left (b^2 + \frac{2}{3}bf + \frac{1}{5}f^2\right)P_{mm}(k)\,,
\end{equation}
\begin{equation}
    \label{gg2}
    P^2_{gg}(k) = \left (\frac{4}{3}bf + \frac{4}{7}f^2\right)P_{mm}(k)\,,
\end{equation}
\begin{equation}
    \label{gg4}
    P^4_{gg}(k) = \frac{8}{35}f^2\,P_{mm}(k)\,,
\end{equation}
\begin{equation}
    P^0_{g\theta}(k) = \left (bf + \frac{1}{3}f^2\right)P_{mm}(k)\,,
\end{equation}
\begin{equation}
    P^2_{g\theta}(k) = \frac{2}{3}f^2\,P_{mm}(k)\,,
\end{equation}
\begin{equation}
    P^0_{\theta\theta}(k) = f^2\,P_{mm}(k)\,.
\end{equation}
The power spectrum P$_{gg}(\vec{k})$ can be written as a multipole expansion as a function of angle to the local line of sight with direction $\hat{x}$,
\begin{equation}
    \label{ggexpan}
    P_{gg}(\vec{k}) = \sum_{\ell}P^{\ell}_{gg}(k)L_{\ell}(\hat{k}.\hat{x})\,,
\end{equation}
where $P^{\ell}_{gg}(k)$ are the galaxy power spectrum multipoles (equations \ref{gg0}, \ref{gg2}, \ref{gg4} ).

RSD also causes non-linear damping of these power spectra. For the galaxy density power spectrum we model this using a factor:
\begin{equation}
    D_g^2(k , \mu) = \frac{1}{1 + (k\mu \sigma_v / H_0)^2}\,,
\end{equation}
where $H_0 = 100 \, h$ km s$^{-1}$ Mpc$^{-1}$ and $\sigma_v$ is the non-linear velocity dispersion parameter in units of km s$^{-1}$. For the velocity power spectrum we also include an additional term in our correlation function models to account for the strength of damping, in the form of a sinc function introduced by \citet{Koda2014},
\begin{equation}
    D_u(k, \sigma_u) = \frac{\sin(k\sigma_u)}{k\sigma_u}\,,
\end{equation}
where we take $\sigma_u = 13.0 \, h^{-1}$ Mpc, as this is the preferred value of \citet{Koda2014}.
Modifying equations \ref{ggpower}, \ref{gthpower}, \ref{ththpower} to account for damping effects we obtain:
\begin{equation}
    P_{gg}(k , \mu) = (b + f\mu^2)^2\,D^2_g(k,\mu)\,P_{mm}(k)\,,
\end{equation}
\begin{equation}
    P_{g\theta}(k , \mu) = (b + f\mu^2)f\,D_g(k,\mu)\,D_u(k, \sigma_u)\,P_{mm}(k)\,,
\end{equation}
\begin{equation}
    P_{\theta\theta}(k , \mu) = f^2\,D^2_u(k, \sigma_u)\,P_{mm}(k)\,.
\end{equation}
We can find the non-linear power spectrum multipoles by numerically integrating these functions multiplied by the Legendre polynomials. We note that, in our linear-theory model for large-scale fits, we neglect the difference between the momentum power spectrum traced by galaxies and the velocity power spectrum, although this may be modelled via further non-linear corrections as discussed by \cite{Howlett2017}.

\subsection{Galaxy auto-correlation function multipoles}
\label{sec:ggauto}
The auto-correlation function between galaxy overdensity $\delta_g$ at positions $\vec{x}$ and $\vec{x'}$ is
\begin{equation}
    \label{xiggdef}
    \langle \delta_g(\vec{x}) \delta_g(\vec{x'}) \rangle = \int \frac{d^3 \vec{k}}{(2\pi)^3}P_{gg}(\vec{k})e^{i\vec{k}.\vec{r}}\,,
\end{equation}
where $\vec{r} = \vec{x'} - \vec{x}$. The baryon acoustic oscillation (BAO) peak is not well defined by a linear power spectrum model, and so we modify the power spectrum to damp the BAO peak using the no-wiggles power spectrum model from \citet{Eisenstein1998}.  We choose $\Sigma_{\rm nl} = 10 \, h^{-1}$ Mpc for this analysis, where $\Sigma_{\rm nl}$ is the BAO damping parameter.

Substituting Equation \ref{ggexpan} in Equation \ref{xiggdef}, we can identify the multipoles of the galaxy correlation function in terms of the power spectrum multipoles:
\begin{equation}
    \label{ggmulti}
    \xi^{\ell}_{gg}(r) =  \frac{i^{\ell}}{2\pi^2} \int  dk k^2 j_{\ell}(kr) P^{\ell}_{gg}(k)\,,
\end{equation}
where $j_\ell(x)$ is a spherical Bessel function, from which we can see that $\xi^0_{gg}(r)$, $\xi^2_{gg}(r)$ and $\xi^4_{gg}(r)$ will be non-zero. We provide a full derivation of Equation \ref{ggmulti} in Appendix \ref{app:ggderiv}. 

\subsection{Galaxy-velocity cross-correlation function multipoles}
\label{sec:gvcross}
The cross-correlation function between galaxy overdensity $\delta_g$ at position $\vec{x}$ and radial peculiar velocity $u$ at position $\vec{x'}$ is
\begin{equation}
    \langle \delta_g(\vec{x}) u(\vec{x'}) \rangle = \int \frac{d^3 \vec{k}}{(2\pi)^3} k^{-1} P_{g\theta}(\vec{k})(\hat{k}.\hat{x'})e^{i\vec{k}.\vec{r}}\,,
\end{equation}
where $\vec{r} = \vec{x'} - \vec{x}$ \citep{Adams2017}. In order to make the RSD calculation more tractable we make the local flat-sky approximation, $\hat{k}.\hat{x'} \approx \hat{k}.\hat{x}$ for any pair of points. We refer readers to \cite{Castorina2018} or \citet{Dam2021} for curved-sky treatments. Similarly to Equation \ref{ggexpan}, we can write the cross-power spectrum $P_{g\theta}(\vec{k})$ as a multipole expansion as a function of angle to the local line of sight with direction $\hat{x}$ ,
\begin{equation}
    P_{g\theta}(\vec{k}) = \sum_{\ell}P^{\ell}_{g\theta}(k)L_{\ell}(\hat{k}.\hat{x})\,,
\end{equation}
from which we find that $\xi^1_{gu}(r)$ and $\xi^3_{gu}(r)$ are non-zero:
\begin{equation}
    \label{gu1mul}
    \xi^{1}_{gu}(r) = -\frac{aH}{2\pi^2}\int dk k j_{1}(kr)\left(P^{0}_{g\theta}(k) + \frac{2}{5}P^{2}_{g\theta}(k)\right)\,,
\end{equation}
\begin{equation}
    \label{gu3mul}
    \xi^{3}_{gu}(r) = \frac{aH}{2\pi^2}\int dk k j_{3}(kr)\left(\frac{3}{5}P^{2}_{g\theta}(k) + \frac{4}{9}P^{4}_{g\theta}(k)\right)\,.
\end{equation}
See Appendix \ref{app:guderiv} for the full derivation of Equations \ref{gu1mul} and \ref{gu3mul}.

We do not use the galaxy-galaxy hexadecapole and galaxy-velocity octupole in our analysis because there is no detectable signal for these statistics. The 6dFGSv sample is not large enough and velocity errors are too broad to make these statistics worthwhile to employ in this work, this will be worth revisiting in the future when we expect to have access to larger, more accurate velocity samples. 

\subsection{Velocity auto-correlation function multipoles}

We include the velocity auto-correlation information using the standard $\psi_1$ and $\psi_2$ statistics \citep{Gorski1988}.  By considering survey geometry in the models of $\psi_1$ and $\psi_2$ we are able to capture all of the information about linear-theory RSD without needing to perform a multipole expansion. This means the models employed in \citet{Turner2021} can also be utilised here, and we summarise how we arrive at these models.

We first define $\Psi_\parallel(r)$ and $\Psi_\perp(r)$, which are functions describing the correlation between components of velocity parallel and perpendicular to the separation vector $\vec{r}$. The spectral form of $\Psi_\parallel(r)$ and $\Psi_\perp (r)$ was described by \citet{Gorski1988},
\begin{equation}
    \label{psipara}
    \Psi_\parallel(r) = \frac{H^2 a^2 (f\sigma_8)^2}{2\pi^2} \int \frac{P_{mm}(k)}{\sigma_{8,{\rm fid}}^2} \left[ j_0(kr) - 2\frac{j_1(kr)}{kr} \right] dk\,,
\end{equation}
\begin{equation}
    \label{psiperp}
    \Psi_\perp(r) = \frac{H^2 a^2 (f\sigma_8)^2}{2\pi^2} \int \frac{P_{mm}(k)}{\sigma_{8,{\rm fid}}^2} \frac{j_1(kr)}{kr} dk\,.
\end{equation}

The models for $\psi_1$ and $\psi_2$ can be expressed as a function of both $\Psi_\parallel(r)$ and $\Psi_\perp(r)$,
\begin{equation}
    \langle \psi_{1}(r) \rangle = \mathcal{A}(r)\Psi_\parallel(r) + [1 - \mathcal{A}(r)]\Psi_\perp(r)\,,
\end{equation}
\begin{equation}
    \langle \psi_{2}(r) \rangle = \mathcal{B}(r)\Psi_\parallel(r) + [1 - \mathcal{B}(r)]\Psi_\perp(r)\,,
\end{equation}
where $\mathcal{A}$ and $\mathcal{B}$ are functions describing the geometry of the survey, dictating the contributions of $\Psi_\parallel$ and $\Psi_\perp$ to $\psi_1$ and $\psi_2$, respectively,
\begin{equation}
    \mathcal{A}(r) = \frac{\Sigma \,w_A w_B\, \cos{\theta_A} \cos{\theta_{B}} \cos{\theta_{AB}}}{\Sigma \,w_A w_B\, \cos^2{\theta_{AB}}}\,,
\end{equation}
\begin{equation}
    \mathcal{B}(r) = \frac{\Sigma \,w_A w_B\, \cos^2{\theta_A} \cos^2{\theta_{B}}}{\Sigma \,w_A w_B\, \cos{\theta_A} \cos{\theta_{B}} \cos{\theta_{AB}}}\,,
\end{equation}
where the sums are taken over pairs of objects $A$ and $B$ in each separation bin, where $w_{A,B}$ are the weights of each object, and $\theta_A$, $\theta_B$ and $\theta_{AB}$ define the geometry of the pair with respect to the observer, following Figure \ref{fig:geom}.

\section{Methodology}
\label{sec:estimators}
\subsection{Estimators}

In this section, we will describe how we build estimators for the galaxy and velocity correlations. The non-Gaussian error distribution in the 6dFGS velocity data forces us to move to the Gaussian-distributed logarithmic distance ratio parameter as the observable instead. To accommodate this we need to rebuild our estimators and optimal FKP weights for the galaxy and velocity correlations with the $\eta$ variable in mind, while also incorporating the effects of linear RSD.

In \citet{Turner2021} we studied estimators for galaxy and velocity correlations for the velocity observable, in real space. In this case we considered only the magnitude of separation between a galaxy pair in the construction of our estimators, $s$. We now extend this construction to also consider $\mu$, binning pair counts in $(s, \mu)$ space to model the RSD effects which depend on the angle to the line of sight. For our cross-correlation multipoles, $\mu$ can take values from $-1$ to $1$, and for our galaxy auto-correlation multipoles, $\mu$ can take values from $0$ to $1$. We bin $\mu$ in steps of 0.1 between these two sets of bounds.

We introduce an FKP weighting \citep{Feldman1994} for our galaxy density sample,
\begin{equation}
    W_{g} = \frac{1}{P_{gg}\,n_i + 1}\,,
    \label{eq:ggfkp}
\end{equation}
where $P_{gg} = 10^4 \, h^{-3}$ Mpc$^3$ is the characteristic amplitude of the galaxy power spectrum and $n_i$ is the number density of the redshift sample at galaxy $i$. 

The $(s, \mu)$ 6dFGS data can be passed to the longform estimators discussed in \citet{Turner2021}, with modifications made to the weights in order to correctly compute the density-$\eta$ pair counts.
The optimal weight to be applied to the $\eta$ pair counts in order to produce an unbiased measurement of the correlation function is the revised FKP weighting, $W_i = w_i / \alpha_i$, where $\alpha_i$ is the conversion factor defined in Equation \ref{eqalpha} and $w_i$ is the optimal FKP weighting applied to a velocity sample,
\begin{equation}
    \label{eq:newFKP}
    W_i = \frac{1}{\sigma_{\eta,i}^2/\alpha_i + \, \alpha_i \,
    n_i \, P_v}\,,
\end{equation}
where $P_v = 10^9 \, h^{-3}$ Mpc$^3$ km$^2$ s$^{-2}$ is the characteristic amplitude of the velocity power spectrum and $n_i$ is the number density of the velocity sample at galaxy $i$. Following \citet{Qin2019}, we check the dependence of our results on our choices for $P_{gg}$ and $P_v$ and find that they are insensitive to the value of FKP amplitudes, in agreement with the findings from that work.

The normalised estimator for the velocity auto-correlation function is:
\begin{equation}
  \hat{\xi}_{vv} = N^2 \, \frac{D_vD_v}{R_gR_g} - 2 N \,
  \frac{D_vR_v}{R_gR_g} + \frac{R_vR_v}{R_gR_g}\,,
\label{eqautoest}
\end{equation}
where $N$ is the normalisation constant,
\begin{equation}
  N = \frac{\sum_i^{n_R} W^R_i}{\sum_i^{n_D} W^D_i} \approx \frac{n_R}{n_D}
  \label{eq:vvN}\,,
\end{equation}
where there are $n_D$ data objects with weights $W^D_i$ and $n_R$
random objects with weights $W^R_i$, both drawn from Equation \ref{eq:newFKP}. 
We find that the estimator
for the velocity auto-correlation function using $\eta$ pair counts
has the same form as Equation \ref{eqautoest}, with an additional scaling
factor:
\begin{equation}
    \label{eq:etaVV}
  \hat{\xi}_{vv} = \frac{\langle w^2 \rangle}{\langle w^2 \cdot \alpha^2
    \rangle} \times \left( N^2 \, \frac{D_\eta D_\eta}{R_gR_g} - 2 N
  \, \frac{D_\eta R_\eta}{R_gR_g} + \frac{R_\eta R_\eta}{R_gR_g}
  \right)\,.
\end{equation}
The scaling factor at the front of this equation depends on the
separation bin:
\begin{equation}
    {\frac{\langle w^2 \rangle}{\langle w^2 \cdot \alpha^2 \rangle} =
    \frac{\sum_{\rm data \; pairs \; in \; bin} w^D_i \,
      w^D_j}{\sum_{\rm data \; pairs \; in \; bin} w^D_i \, w^D_j \,
      \alpha^D_i \, \alpha^D_j}} .
\end{equation}
The modifications for $\psi_1$ and $\psi_2$ are analogous to this.

We use a similar formalism for the galaxy-galaxy two-point auto-correlation function, sans the leading scaling factor as there are no velocity terms to convert to terms in logarithmic distance,
\begin{equation}
  \hat{\xi}_{gg} = N^2 \, \frac{D_gD_g}{R_gR_g} - 2N \,
  \frac{D_gR_g}{R_gR_g} + 1 + IC\,.
  \label{eq:ggautoest}
\end{equation}
$N$ has identical formalism to Equation \ref{eq:vvN}, but the optimal weights are FKP weights tailored for the galaxy density sample given by Equation \ref{eq:ggfkp}.  The term $IC$ in Equation \ref{eq:ggautoest} refers to the integral constraint correction \citep{Peebles1974b, Peebles1980, Landy1993, Scranton2002}, a factor used to correct for the additive bias introduced to the estimate due to the fact that we cannot measure the true number density of galaxies from our catalogues. We can determine the integral constraint as,
\begin{equation}
    IC = \frac{\sum_i \xi_{gg}(r_i) \, R_gR_g(r_i)}{\sum_i R_gR_g(r_i)}\,,
\end{equation}
where $\xi_{gg}$ is the best-fitting galaxy-galaxy correlation function monopole, and $R_gR_g(s)$ is the random-random galaxy pair count. For the 6dFGS sample we derive $IC = 5.92\times10^{-4}$.

Generalising the formalism used to create the velocity auto-correlation function estimator for the cross-correlation, we can write the normalised estimator in terms of the velocity variable as,
\begin{equation}
    \begin{aligned}
    \hat{\xi}_{gv} = &\left( \frac{\sum_i^{n_{R1}} w^{R1}_i
    \sum_i^{n_{R2}} w^{R2}_i}{\sum_i^{n_{D1}} w^{D1}_i \sum_i^{n_{D2}}
    w^{D2}_i} \right) \frac{D_{1g}D_{2v}}{R_{1g}R_{2g}} - 2\left(
    \frac{\sum_i^{n_{R1}} w^{R1}_i}{\sum_i^{n_{D1}} w^{D1}_i} \right)
    \frac{D_{1g}R_{2v}}{R_{1g}R_{2g}} \\ 
    &- 2\left(
    \frac{\sum_i^{n_{R2}} w^{R2}_i}{\sum_i^{n_{D2}} w^{D2}_i} \right)
    \frac{R_{1g}D_{2v}}{R_{1g}R_{2g}} +
    \frac{R_{1g}R_{2v}}{R_{1g}R_{2g}}\,,
    \end{aligned}
    \label{eqcrossest}
\end{equation}
and by translating the equation from velocity to $\eta$ we arrive at an equation with a similar formalism to Equation \ref{eq:etaVV}, where the scaling factor at the front of this equation also depends on the separation bin,
\begin{equation}
    \begin{aligned}
      \hat{\xi}_{gv}' &= \frac{\langle w^2 \rangle}{\langle w^2 \cdot \alpha \rangle} \times \hat{\xi}_{g\eta} \\&= \frac{\sum_{\rm data \; pairs \; in \; bin} w^{D1} \,
      w^{D2}}{\sum_{\rm data \; pairs \; in \; bin} w^{D1} \, w^{D2} \,
      \alpha^{D2}} \times \hat{\xi}_{g\eta}\,.
    \end{aligned}
\end{equation}

\subsection{Fitting}
\label{ssec:fits}
We calculate the multipole moments of our correlation function measurements, guided by the non-zero multipoles of our models:
\begin{equation}
    \label{multipolesum}
    \xi^{\ell}(s) = \frac{2\ell + 1}{2}\int d\mu \, \xi(s, \mu) \, L_{\ell}(\mu)\,,
\end{equation}
where $\xi(s, \mu)$ are our correlation function measurements, $L_{\ell}(\mu)$ are Legendre polynomials in terms of $\mu$, and d$\mu = 0.1$. Using Equation \ref{multipolesum} to integrate over $\mu$, the 6dFGS data binned in $(s, \mu)$ can be collapsed to a one-dimensional vector binned by the magnitude of the separation between galaxy pairs, $s$. The same equations are used to compute the multipoles of the 6dFGS mocks and the 6dFGS data. We use five correlation functions in our fit: the galaxy-galaxy monopole and quadrupole $\xi_{gg}^0$ and $\xi_{gg}^2$, the cross-correlation dipole $\xi_{gu}^1$, and the $\psi_1$ and $\psi_2$ velocity auto-correlation statistics. We compute the mock mean and standard deviation values across all 600 6dFGS mocks, from which we construct the covariance matrix of our measurements.

We opt to perform our parameter search over a pre-computed grid rather than using an MCMC algorithm. We calculate the values of our pre-computed multipole models for values of $\sigma_v$ between $0$ km s$^{-1}$ and $550$ km s$^{-1}$ in steps of $50$ km s$^{-1}$, and values of $\beta$ between $0.00$ and $4.00$ in steps of $0.10$. We establish a uniform prior on $f$ and $b$ within the ranges $0.00 < f < 1.25$ and $0.50 < b < 2.20$. We use a step-size $\Delta f = \Delta b = 0.025$ for the fits to the mock datasets, and a finer step-size of $\Delta f = \Delta b = 0.0025$ for the fits to the 6dFGS survey data. Iterating through these ranges in order to compute $\beta$, we then construct the corresponding multipole models using this value of $\beta$ by linearly interpolating between the closest two values of $\beta$ from the pre-computed grid.

We do this for all combinations of $f$ and $b$, whilst also iterating over the 12 values of $\sigma_v$ from our grid, performing the same chi-squared minimisation procedure from \citet{Turner2021} to find the combination of parameters ($f$, $b$, $\sigma_v$) that minimises the equation
\begin{equation}
    \label{eq:chisq}
    \begin{aligned}
    \chi^2(f, b, \sigma_v) = \sum_{i,j = 1}^N ~& (A_d(i) - A_m(i; f, b, \sigma_v)) ~\hat{C}_{ij}^{-1}\\
    & (A_d(j) - A_m(j; f, b, \sigma_v))\,,
    \end{aligned}
\end{equation}
where $\hat{C}_{ij}$ is the covariance matrix constructed from the 600 6dFGS mocks and A$_{d,m}$ are the concatenated vectors of estimator measurements for the data/mocks and the models, respectively. The limited number of mocks used to construct the covariance matrix will introduce additional noise to our analysis upon inverting the matrix, leading to an underestimate of the size of our confidence regions if not accounted for. We mitigate this by including the corrective factor for the inverse covariance matrix from \citet{Hartlap2007}:
\begin{equation}
    \hat{C}^{-1} = \frac{n - p - 2}{n - 2}C^{-1}\,,
\end{equation}
where the number of mocks $n = 600$, the number of entries in our data vector $p = 79$, and $C^{-1}$ is our original, uncorrected inverse covariance matrix. Applying this term lowers our $\chi^2$ values and marginally broadens our uncertainties, but does not impact our mean results.

\begin{figure*}
    \centering
    \includegraphics[width = 0.8\textwidth]{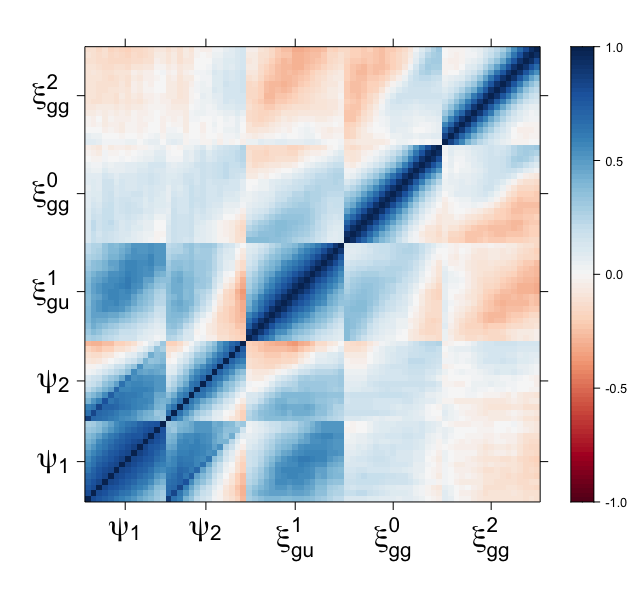}
    \caption{Correlation matrix, dimensions 79 x 79, for the full concatenated data vector of all five statistics used in our final 6dFGS analysis. Each cell corresponds to a separation bin within the chosen fitting range, where we use $21 < s < 99 \, h^{-1}$ Mpc for $\psi_1$ and $\psi_2$, and $27 < s < 123 \, h^{-1}$ Mpc for the three multipoles, both with bin width $6 \, h^{-1}$ Mpc.}
    \label{fig:redcovar}
\end{figure*}

We show the correlation matrix used to produce our reported result in Figure \ref{fig:redcovar}. Measuring from bin mid-points, we use a fitting range of $21 < s < 99 \, h^{-1}$ Mpc for the velocity-velocity measurements and $27 < s < 123 \, h^{-1}$ Mpc for the multipole measurements, with separation bin width $\Delta s = 6 \, h^{-1}$ Mpc our correlation matrix has dimensions 79x79. The choice of fitting range for velocity measurements is forced by the dimensions of the survey, as the data becomes noisy at large scales as we approach the redshift limit. However our method is robust against the choice of fitting range used for the multipoles, this is discussed further in Section \ref{sec:results}.

\section{Results}
\label{sec:results}
We perform growth rate fits for all 600 6dFGS mocks ("mock results") and for the 6dFGSv and 6dFGSz data catalogues ("6dFGS results") and convert the best-fitting values of $f$ and $b$, as determined from the peaks of the posterior distributions, into results in terms of $f$ and $\beta = f/b$.

\subsection{6dF mock results}
\subsubsection{Full five-statistic fit}
\label{sssec:fivemod}
We first show that the estimators we've presented produce an unbiased estimate of the growth rate of structure at the level of statistical accuracy appropriate for our dataset, fitting to all five correlation functions. By taking the mean of the $f\sigma_8$ posterior from all 600 mocks, setting $\sigma_8(z = 0.10) = 0.78$ and taking the ensemble mean of all 600 values, we find $f\sigma_8 = 0.391 \pm 0.079$ and $\beta = 0.394 \pm 0.087$. We do not expect that making this choice of fiducial $\sigma_8$ will impact our best-fitting values at linear order, but note that corrections to the covariance matrix may need to be considered as a result of such a choice \citep[see][]{Hollinger2021}. From the 6dFGS mocks, we recover the growth rate with $20.2$ per cent statistical accuracy and $\beta$ with $22.1$ per cent statistical accuracy. The mean chi-squared value for this fit is 67.3, with 76 degrees of freedom. The peak of the distribution of best-fitting $f\sigma_8$ values underestimates the fiducial cosmology, $f\sigma_{8, \rm fid}(z = 0.1) = 0.445$, but is nonetheless recovered to within $1\sigma$ of the distribution.

\subsubsection{Galaxy-galaxy multipoles}
We also quantify the improvement gained by the inclusion of velocity correlations by applying the method described in Section \ref{sec:estimators} to only the multipoles of the galaxy auto-correlation function. Performing the same analysis as in Section \ref{sssec:fivemod}, limiting our fit to the $\xi_{gg}^0$ and $\xi_{gg}^2$ statistics only, we find $f\sigma_8 = 0.391 \pm 0.113$ and $\beta = 0.390 \pm 0.120$. In this scenario we recover $f\sigma_8$ with $28.9$ per cent statistical accuracy and $\beta$ with $30.8$ per cent statistical accuracy. The mean chi-squared value for this fit is 30.4, with 31 degrees of freedom. By including the velocity data via the velocity auto-correlation function and velocity-density cross-correlation function multipoles we see a relative improvement of $30.1$ per cent in our estimation of the error in measurements of $f\sigma_8$, and of $27.5$ per cent in measurements of $\beta$, as compared to the accuracy of our results when fitting to the RSD data only.

\subsubsection{Velocity-velocity estimators and galaxy-velocity dipole}
By only considering the galaxy-velocity cross-correlation dipole $\xi_{gu}^1$ and the two velocity-velocity auto-correlation function models $\psi_1$ and $\psi_2$ we can isolate the growth constraints arising from direct peculiar velocity information in our mocks. Following the same steps as in the above cases we find $f\sigma_8 = 0.374 \pm 0.117$ and $\beta = 0.327 \pm 0.138$, recovering cosmological parameters with $31.3$ per cent and $42.2$ per cent accuracy respectively. The mean chi-squared value for this fit is 40.0, with 42 degrees of freedom. The inclusion of information from RSD through the galaxy-galaxy multipoles gives us relative improvements of $32.5$ per cent and $37.0$ per cent in the estimated error on our measurements of $f\sigma_8$ and $\beta$, respectively. 

\begin{figure}
    \centering
    \includegraphics[width = 0.99\columnwidth]{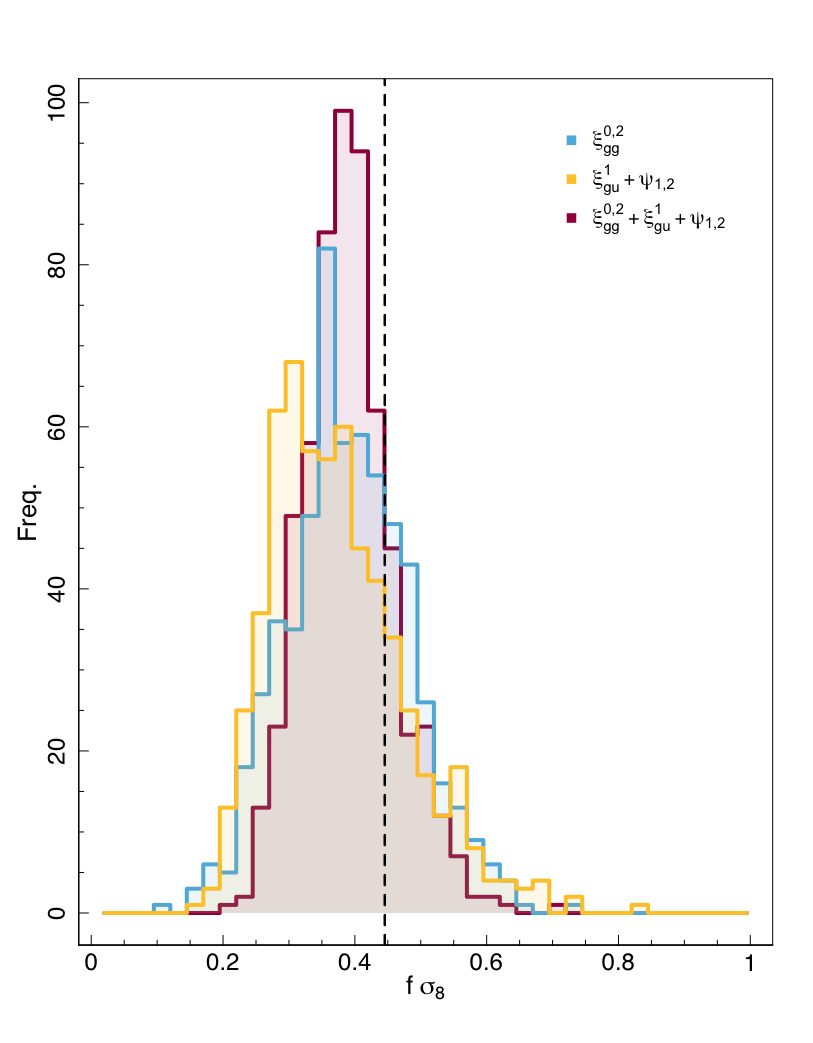}
    \caption{Distributions of $f\sigma_8$ values calculated from posteriors of three different combinations of statistics for 600 6dFGS mocks. The distribution of values calculated using only the galaxy-galaxy multipoles is shown in blue, the distribution of values calculated using the $\psi$ estimators and cross-correlation dipole is shown in yellow, and the results obtained from the combination of these two groups -- the full five-statistic fit -- is given by the purple distribution.}
    \label{fig:minhist}
\end{figure}

We summarise this comparative analysis in Figure \ref{fig:minhist}, in which we show the distributions of 600 $f\sigma_8$ values from each mock, calculated from the mean of the growth rate posterior for these three combinations of statistics. The distribution of 600 galaxy-galaxy multipoles only fits is shown in blue, the distribution of 600 results when fitting to the $\psi$ estimators and cross-correlation dipole is shown in yellow, and the distribution of the fits to the combination of all five statistics is shown in purple. The fiducial value of $f\sigma_{8, \rm fid}(z = 0.10) = 0.445$ for the 6dFGS mocks is shown by the dashed black line, all three treatments covered recover the fiducial cosmology to within $1\sigma$. 

\subsection{6dF data results}

Before presenting our fiducial growth rate fits to the 6dFGS data, we first verified that our results were not significantly dependent on the fitting range employed. To accomplish this we fit to the data using several different reasonable considerations of the fitting range. We choose a set of three bin mid-points to act as the lower bound of the range, $s_{\rm min} = 21, 27, 33 \, h^{-1}$ Mpc, and another set of five to act as upper bounds, $s_{\rm max} = 99, 111, 123, 135, 147 \, h^{-1}$ Mpc. The choice of minimum points is motivated by the applicability of our theoretical model, excluding small-scale non-linearity that we do not account for, and the choice of maximum points is motivated by the dimensions of the 6dFGSv dataset as defined by the effective redshift of the sample. This leaves us with a set of 15 total fitting ranges to test, covering a minimum range of $66 \, h^{-1}$ Mpc and a maximum range of $126 \, h^{-1}$ Mpc. If the results of our chi-squared analysis were dependent on the choice of fitting range it would be seen in a comparative analysis of all 15 tests.

\begin{figure*}
    \centering
    \includegraphics[width = \textwidth]{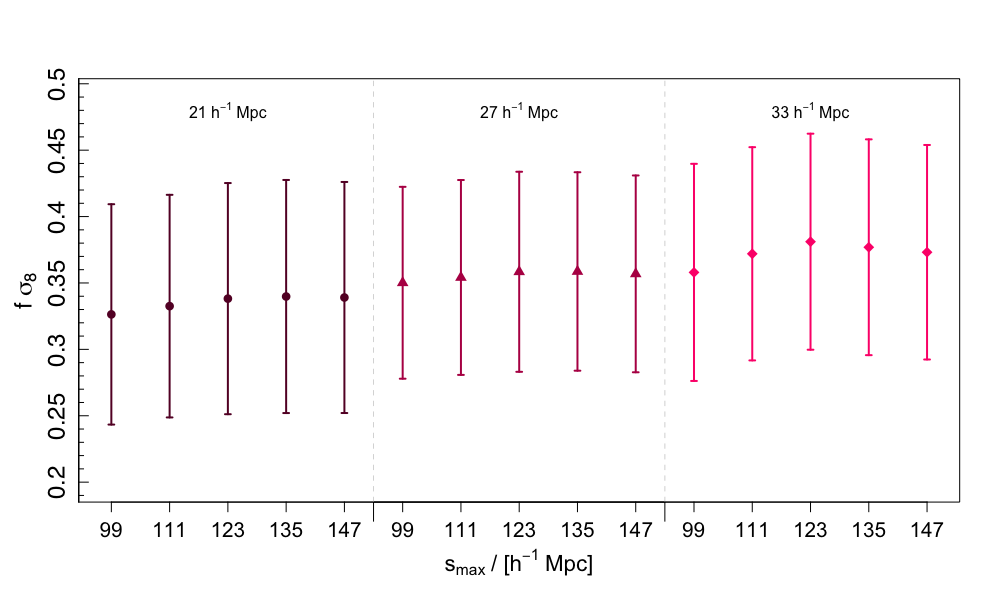}
    \caption{Values of the normalised growth rate, $f\sigma_8$ measured from the marginalised posteriors of five-statistic fits to the 6dFGS data using different fitting ranges. Ranges are sorted into three blocks by $s_{\rm min}$ values; $21 \, h^{-1}$ Mpc (circles), $27 \, h^{-1}$ Mpc (triangles) and $33 \, h^{-1}$ Mpc (diamonds), and are ordered in terms of $s_{\rm max}$ values as given on the x-axis.}
    \label{fig:fitrange}
\end{figure*}

In Figure \ref{fig:fitrange} we show the recovered values of $f$ for the set of 15 fitting ranges. The recovered values of $f$ shown are the means and $1\sigma$ errors measured from the $f$ posteriors.  From this analysis we can see that, while the value of $f$ does systematically increase as we increase the size of the fitting range, the errors we obtain on each measurement are comparatively independent of our choice and all of the values of the growth rate that we obtain agree with one another within $1\sigma$. Assuming $\sigma_8(z = 0.0) = 0.82$, the minimum value we find is $f\sigma_8 = 0.326 \pm 0.083$ when fitting between $21 < s < 99 \, h^{-1}$ Mpc and the maximum is $f\sigma_8 = 0.381 \pm 0.081$ when fitting between $33 < s < 123 \, h^{-1}$ Mpc. We also find $\beta = 0.284 \pm 0.074$ and $\beta = 0.309 \pm 0.071$ at these fitting ranges, respectively. The full set of measurements of $f$ and $b$ are given in Table \ref{tab:fits}. Given the robustness of our method against the choice of $s_{\rm min}$ and $s_{\rm max}$ we select $27 < s < 123 \, h^{-1}$ Mpc as our fitting range. 

\begin{table*}
    \centering
    \begin{tabular}{c|c|c|c|c|c|c|c|c|c}
    \hline
    $s_{\rm min}$ / ($h^{-1}$ Mpc) & $s_{\rm max}$ / ($h^{-1}$ Mpc) & $f$ & $\Delta f$ & $b$ & $\Delta b$ & $\sigma_v$ / (km s$^{-1}$) & $\Delta \sigma_v$ / (km s$^{-1}$) & $\chi^2_\nu$ & $\nu$\\
    \hline
    27 & 123 & 0.437 & 0.092 & 1.464 & 0.089 & 199.6 & 140.1 & 0.76 & 76\\
       &     &       &       &       &       &       &       &      &   \\ 
    21 & 99  & 0.398 & 0.101 & 1.400 & 0.067 & 260.1 & 145.2 & 0.81 & 67\\
    21 & 111 & 0.406 & 0.102 & 1.393 & 0.067 & 272.5 & 146.0 & 0.78 & 73\\
    21 & 123 & 0.412 & 0.106 & 1.396 & 0.067 & 291.2 & 146.3 & 0.77 & 79\\
    21 & 135 & 0.414 & 0.107 & 1.396 & 0.067 & 299.3 & 146.5 & 0.81 & 85\\
    21 & 147 & 0.413 & 0.106 & 1.396 & 0.067 & 295.3 & 146.5 & 0.78 & 91\\
    
    27 & 99  & 0.427 & 0.088 & 1.485 & 0.087 & 179.8 & 132.6 & 0.79 & 64\\
    27 & 111 & 0.432 & 0.089 & 1.473 & 0.088 & 191.8 & 137.0 & 0.76 & 70\\
    27 & 135 & 0.437 & 0.091 & 1.479 & 0.087 & 195.9 & 138.6 & 0.79 & 82\\
    27 & 147 & 0.435 & 0.090 & 1.482 & 0.087 & 192.3 & 137.3 & 0.76 & 88\\
    
    33 & 99  & 0.435 & 0.100 & 1.536 & 0.119 & 181.0 & 136.9 & 0.82 & 61\\
    33 & 111 & 0.454 & 0.098 & 1.517 & 0.120 & 181.1 & 137.3 & 0.78 & 67\\
    33 & 123 & 0.465 & 0.099 & 1.504 & 0.122 & 183.7 & 139.1 & 0.78 & 73\\
    33 & 135 & 0.460 & 0.099 & 1.524 & 0.120 & 184.6 & 139.2 & 0.81 & 79\\
    33 & 147 & 0.455 & 0.099 & 1.535 & 0.119 & 183.2 & 138.4 & 0.78 & 85\\
    \hline
    \end{tabular}
    \caption{Measurements and errors for the growth rate and linear galaxy bias from the five-statistic fit to the 6dFGS data.  Results are obtained from the one-dimensional marginalised posteriors following the method detailed in Section \ref{ssec:fits}, applying different fitting ranges bounded by $s_{\rm min}$ and $s_{\rm max}$ to the cross-correlation dipole and the galaxy auto-correlation multipoles. The reduced chi-squared value is included to indicate goodness-of-fit. Given the robustness of these results to the choice of $s_{\rm min}$ and $s_{\rm max}$ we present the result obtained from the range $27 < s < 123 \, h^{-1}$ Mpc as our final result, which is given in the first row of the table.}
    \label{tab:fits}
\end{table*}

From this choice of fitting range we recover $f\sigma_8 = 0.358 \pm 0.075$ ($20.9$ per cent uncertainty) and $\beta = 0.298 \pm 0.065$ ($21.8$ per cent uncertainty) from the fit to the five statistics $(\xi_{gg}^0, \xi_{gg}^2, \xi_{gu}^1, \psi_1, \psi_2)$. The chi-squared value for this fit is 57.9, with 76 degrees of freedom.

\subsubsection{Galaxy-galaxy multipoles}
We repeat the RSD-only analysis from Section \ref{sssec:fivemod} with the 6dFGS dataset over the same fitting range, $27 < s < 123 \, h^{-1}$ Mpc. Fitting to the galaxy-auto correlation function multipoles only, we recover $f\sigma_8 = 0.441 \pm 0.108$ ($24.5$ per cent uncertainty) and $\beta = 0.387 \pm 0.100$ ($25.8$ per cent uncertainty). The chi-squared value for this fit is 43.7, with 31 degrees of freedom. We still see an improvement in our uncertainties by including the information contained in the velocity data, similar to that found in the mock analysis. The constraints on our measurement of $f\sigma_8$ and $\beta$ are improved by $30.6$ per cent and $35.0$ per cent, respectively, in the full five-statistic fit by the inclusion of velocity information as opposed to when we fit solely to the galaxy-galaxy multipoles. 

\subsubsection{Velocity-velocity estimators and galaxy-velocity dipole}
Fitting only to the cross-correlation dipole and the velocity-auto-correlation function estimators over the range $27 < s < 123 \, h^{-1}$ Mpc, we recover $f\sigma_8 = 0.249 \pm 0.132$ ($53.0$ per cent uncertainty) and $\beta = 0.245 \pm 0.160$ ($65.3$ per cent uncertainty). The chi-squared value for this fit is 15.4, with 42 degrees of freedom. These measurements and the unexpectedly low $\chi^2$ value for the 6dFGS data PV analysis highlight the limitations of current velocity datasets for this kind of analysis when applied without a density counterpart. Additionally, the 6dFGS mock catalogues are unable to replicate the low mass halos observed in the data at low redshifts, $z < 0.05$, due to the mass resolution used \citep{Carter2018}. It may also be the case that the 6dFGS velocity errors are over-estimated, but this issue is not something we were able to pursue. The results we present when utilising the 6dFGSv dataset should thus be considered conservative, taking these issues into account. It is likely that the 6dFGS velocity data is more constraining than we have assumed.

The constraints on our measurement of $f\sigma_8$ and $\beta$ are improved by $43.2$ per cent and $59.4$ per cent, respectively, in the full five-statistic fit by the inclusion of galaxy information, as opposed to when we fit solely to the cross-correlation dipole and the $\psi$ estimators. 

\begin{figure*}
    \centering
    \includegraphics[width = 0.95\textwidth]{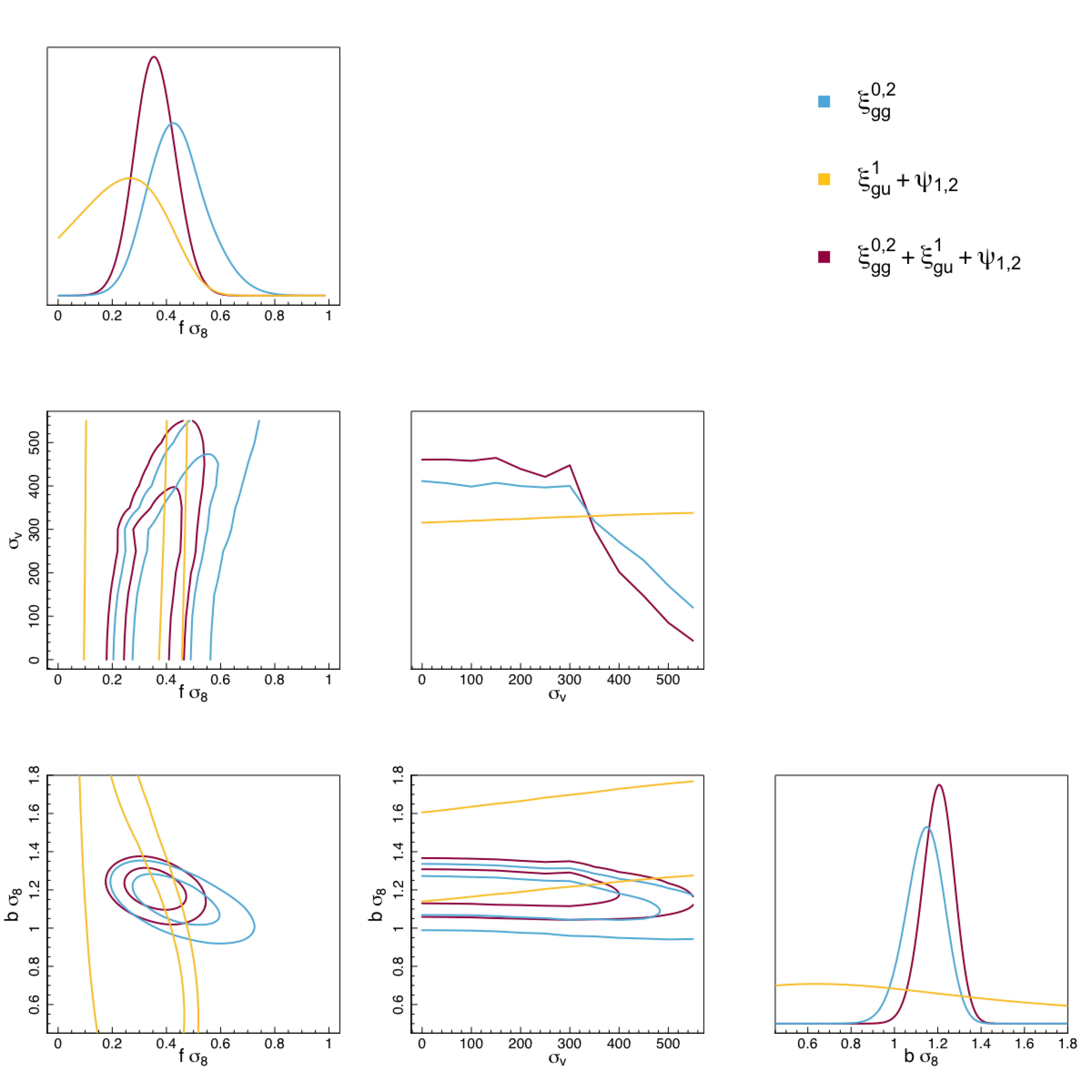}
    \caption{Posteriors of $f\sigma_8$, $b\sigma_8$ and $\sigma_v$ for 6dFGS from various combinations of the statistics we present in Section \ref{sec:models}; the galaxy-galaxy multipoles only ($\xi_{gg}^{0,2}$) in blue, the velocity estimators and cross-correlation dipole ($\psi_{1,2}+\xi_{gu}^1$) in yellow and the full five-statistic fit in purple.}
    \label{fig:contour}
\end{figure*}

The results we obtain by separately applying our method to the 6dFGSv and 6dFGSz samples work together to highlight the benefit of combining these datasets in one larger joint analysis. The improved constraining powers of the joint analysis of the galaxy and velocity correlation function multipoles, as opposed to the individual galaxy and velocity analyses, can be seen both in the 6dFGS mocks in Figure \ref{fig:minhist} and in the 6dFGS data in Figure \ref{fig:contour}.

In Figure \ref{fig:contour} we present the posteriors of $f\sigma_8$, $b\sigma_8$ and $\sigma_v$ for 6dFGS in the range $27 < s < 123 \, h^{-1}$ Mpc. Similarly to Figure \ref{fig:minhist}, we show the fit to the galaxy-galaxy multipoles in blue, the cross-correlation dipole and $\psi$ estimators in yellow, and the combination of all five statistics in purple. From the relatively unbounded nature of the yellow ($\psi_1 + \psi_2 + \xi_{gu}$) dipole contours in Figure \ref{fig:contour}, we can see that the main source of statistical uncertainty in our constraints of $f\sigma_8$ come from the velocity component of the analysis. This is also corroborated by Figure \ref{fig:minhist}, as seen by the broader distribution of mean $f\sigma_8$ values recovered from our 6dFGS mock analysis when fitting to this same combination of statistics, as opposed to either the RSD-only fit or the full five-statistic fit. Improvements to future velocity datasets, in terms of both size and measurement accuracy, will in turn make our measurement of the growth rate more accurate.

\begin{figure*}
    \centering
    \includegraphics[width = 0.8\textwidth]{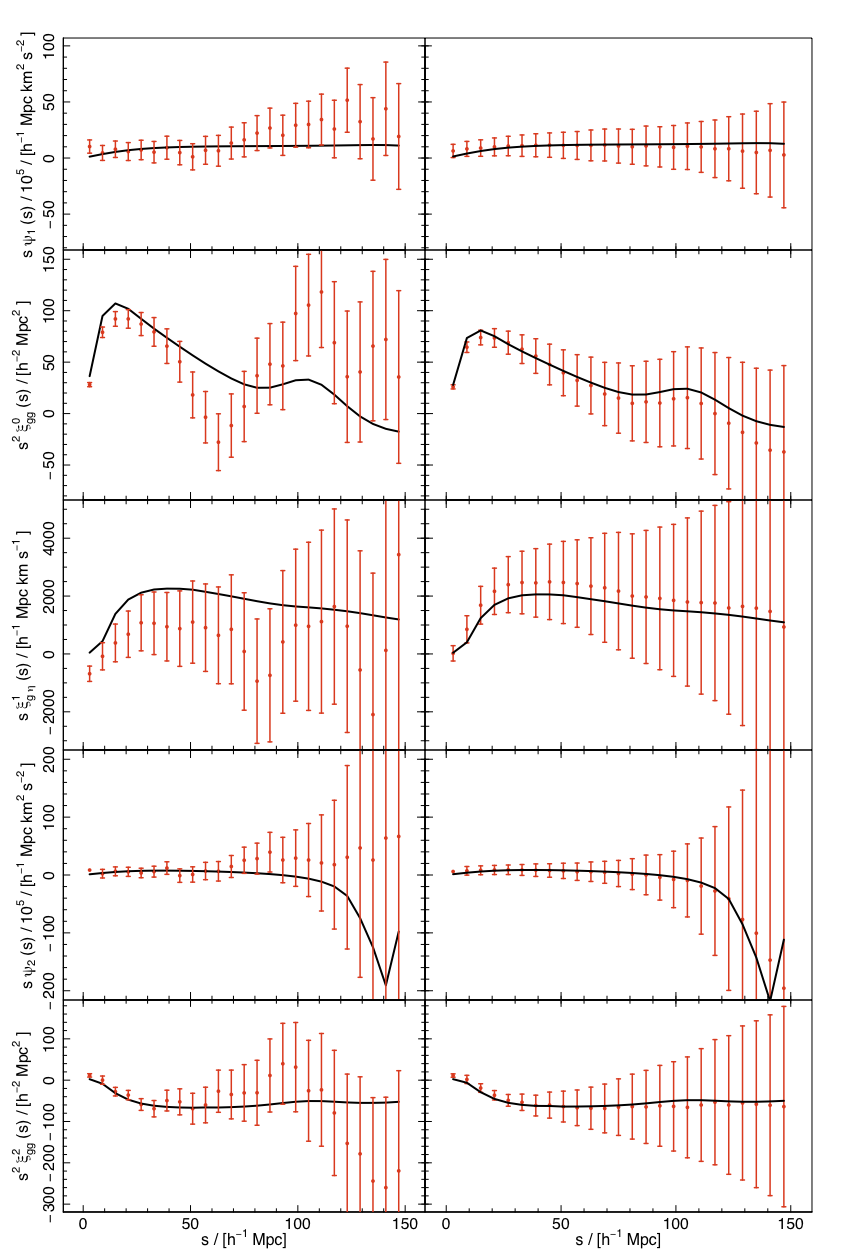}
    \caption{Models of the 5 statistics used in our analysis in black, compared to the measurements of the corresponding estimators from the 6dFGS data in orange-red. Plotted alongside this in the righthand column are the mean measurements of the corresponding estimators from the 600 6dFGS mocks. Errorbars represent the standard deviation in these 600 measurements.}
    \label{fig:model_v_meas}
\end{figure*}

In Figure \ref{fig:model_v_meas} we present our measurements of the five considered estimators, described in Section \ref{sec:estimators}, from the 6dFGSv and 6dFGSz data catalogues (left column) and the 6dFGS mock catalogues (right column). The models as described in Section \ref{sec:models} are shown in black, and the measurements are depicted by the orange errorbars. The errors themselves are the standard deviation in the measurement across all 600 mocks centred on the mean measurement in that separation bin, and this is consistent across all ten sub-figures. In all cases, the mock measurements are able to recover the model to within one standard deviation. Given that the data represents a single possible realisation, the estimators also provide a good approximation of what we would expect from the models.

\section{Comparison with other local growth rate measurements}
\label{sec:6dfcomp}
We now place these measurements into a broader context, comparing our constraints with other recent results. This comparison is split into several categories: those that use 6dFGS data and therefore produce measurements that are not entirely independent of one another, those that do not use 6dFGS data but employ similar methods, and those that reconstruct the peculiar velocity field from the galaxy overdensity field. 

\begin{figure*}
    \centering
    \includegraphics[width = 0.8\textwidth]{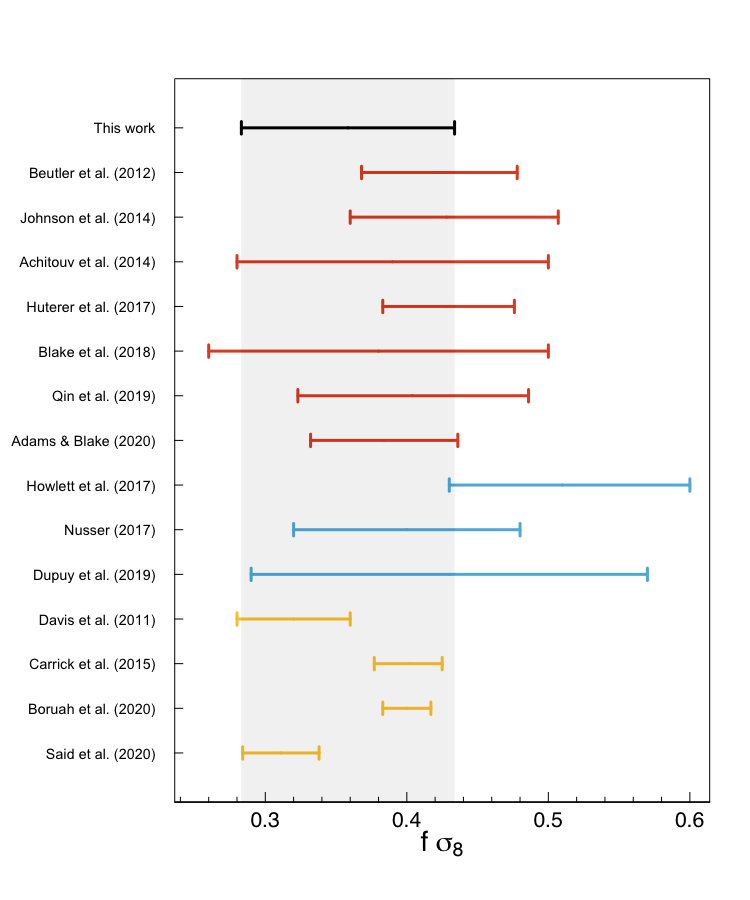}
    \caption{Comparison between the $f\sigma_8$ value presented in this work using the fitting range $27 < s < 123 \, h^{-1}$ Mpc -- also shown as the central errorbar in Figure \ref{fig:fitrange} -- and various reported local measurements of $f\sigma_8$ from contemporary papers. Our result is shown in black, results that also use 6dFGS data are shown in orange-red, results that do not use 6dFGS data are shown in blue, and results that specifically use a velocity field reconstruction technique are shown in yellow. In order, these results are: 1. This work, 2. \citet{Beutler2012}, 3. \citet{Johnson2014}, 4. \citet{Achitouv2017}, 5. \citet{Huterer2017}, 6. \citet{Blake2018}, 7. \citet{Adams2020}, 8. \citet{Qin2019}, 9. \citet{Howlett2017}, 10. \citet{Nusser2017}, 11. \citet{Dupuy2019}, 12. \citet{Said2020}, 13. \citet{Boruah2020}, 14. \citet{Davis2011}, 15. \citet{Carrick2015}}
    \label{fig:6dfcomp}
\end{figure*}

This discussion is summarised in Figure \ref{fig:6dfcomp}: our measurement is shown in black, 6dFGS measurements in orange, non-6dFGS measurements in blue and reconstruction methods in yellow. We have chosen to compare our results to an ensemble of other recent results from the literature that encompass various datasets, methodologies and techniques to measure the local normalised growth rate, and we find that all results are in broad agreement.

\subsection{6dFGS redshift and velocity results}

Our growth rate measurements are broadly consistent with a series of previous 6dFGS analyses.  We briefly summarise each of these studies here, and note that each measurement is displayed in Figure \ref{fig:6dfcomp}.

\citet{Beutler2012} completed the first RSD analysis of the 6dFGS redshift sample, using a K-band selected subsample containing 81 971 galaxies with a median redshift of $z = 0.05$, but modelling the 2D galaxy correlation function at an effective redshift of $z_{\rm eff} = 0.067$. From this RSD analysis, they found $f\sigma_8 = 0.423 \pm 0.055$.

\citet{Johnson2014} found $f\sigma_8 = 0.428^{+0.079}_{-0.068}$ from a velocity auto-covariance analysis of the 6dFGSv dataset using a maximum-likelihood methodology. A similar approach is adopted by \citet{Huterer2017}, using the same 6dFGSv dataset, who find $f\sigma_8 = 0.428^{+0.048}_{-0.045}$. The methods of \citet{Adams2020} are similar to those presented in these papers, using the same 6dFGSv and 6dFGSz samples and extending their analysis to incorporate the effects of RSD. They find $f\sigma_8 = 0.384 \pm 0.052$ (stat.) $ \pm 0.061$ (sys.). We present the statistical error for this result in Figure \ref{fig:6dfcomp}.

\citet{Achitouv2017}, using the 6dFGSz sample and galaxy catalogues used by \citet{Beutler2011} for their BAO analysis, performed a cross-correlation analysis between galaxy overdensity and voids as well as a galaxy auto-correlation analysis, fitting the correlation data to RSD models. In Figure \ref{fig:6dfcomp} we show their measurement of $f\sigma_8 = 0.39 \pm 0.11$, obtained from the galaxy-void cross-correlation, but we note that they also obtain $f\sigma_8 = 0.42 \pm 0.06$ from the galaxy auto-correlation analysis.

\citet{Blake2018} published the first Fourier-space analysis of RSD using the 6dFGS dataset, investigating the Fourier-space multipoles of the power spectrum using the same 6dFGSz sample as \citet{Beutler2011} and \citet{Achitouv2017}. They found $f\sigma_8(z = 0.06) = 0.38 \pm 0.12$. 

\citet{Qin2019} presented a new way of representing the velocity data via the redshift-space momentum power spectrum \citep{Howlett2019}, casting the momentum field as a density-weighted velocity field. They applied the momentum power spectrum, and the redshift-space density power spectrum, to the 6dFGSv sample and the 2MASS Tully-Fisher \citep[2MTF;][]{Masters2008, Hong2014} sample. We show the result they obtain from the 2MTF + 6dFGSv momentum-density cross analysis, $f\sigma_8(z_{\rm eff} = 0.03) = 0.404^{+0.082}_{-0.081}$ in Figure \ref{fig:6dfcomp}, but note that they also presented results from the momentum power spectrum only of $f\sigma_8(z_{\rm eff} = 0.03) = 0.451^{+0.108}_{-0.092}$.

We report a lower, but broadly consistent, measurement of $f\sigma_8$ than other methods that utilise the 6dFGS data. We also implement FKP weighting in our method which reduces our statistical error in relation to other measurements that do not employ such a weighting scheme.  Although all these studies analyse (subsets of) the same dataset, they can be distinguished through using different statistics, scales, configuration space or Fourier space, weighting, and models.  Despite many methodological differences, it is encouraging that all analyses report broadly-consistent results.

\subsection{Other redshift and velocity results}
Despite using different methodologies, the results that we have compared to so far are not independent of one another given that they have all used the 6dFGSv sample, the 6dFGSz sample, or both. It is important to also compare our results with those who have used datasets entirely divorced from 6dFGS as an additional point of reference.

\citet{Howlett2017} present a measurement of the velocity power spectrum, including non-linear RSD, using the peculiar motions of a sample of 2062 galaxies from the 2MTF survey. Assuming a scale-independent behaviour of the growth rate, they found $f\sigma_8 = 0.51^{+0.09}_{-0.08}$. 

\citet{Nusser2017} measured the cross-correlation between radial peculiar velocities from the \textit{cosmicflows-3} \citep[CF3][]{Tully2016} dataset with the dipole moment of the 2MRS galaxy distribution, finding $f\sigma_8 = 0.40 \pm 0.08$.

\cite{Dupuy2019} applied the pairwise velocity estimator $v_{12}$ and $\psi_1$, the same estimator of the velocity auto-correlation function that we employ, to the same CF3 data as \citet{Nusser2017} to produce a local measurement of the normalised growth rate at $z = 0.05$, finding $f\sigma_8 = 0.43 \pm 0.03$ (obs.) $\pm 0.11$ (cosmic).

We note that all these growth rate determinations are also consistent with our result, within the statistical errors.  These measurements are also indicated on Figure \ref{fig:6dfcomp}.

\subsection{Density-velocity reconstruction results}
We have investigated the use of the combined density and velocity fields to produce a measurement of the normalised growth rate. Methods of comparing the measured peculiar velocity field with a modelled peculiar velocity field reconstructed from the observed galaxy overdensity field thus provide a reasonable point of comparison. These methods allow for constraints to be placed on $\beta$, which can be converted to constraints on $f\sigma_8$ following the equation
\begin{equation}
    f\sigma_8 = \beta\sigma_8^g
\end{equation}
where $\sigma_8^g = b\sigma_8$ is the rms fluctuation in the number of galaxies within spheres of radius $8 \, h^{-1}$ Mpc. Thus if one knows the galaxy bias $b$ of the dataset then the growth rate can be recovered. However, as $\sigma_8^g$ is measured directly from the galaxy distribution, it will inevitably include non-linear growth of structure. $\sigma_8$ is defined relative to the amplitude of the linear power spectrum and so a correction for late-time evolution, such as the one put forth by \citet{juszkiewicz2010}, is required. In the following comparisons we quote the linearised value of $f\sigma_{8,lin}$ presented by each work, where possible.

\citet{Davis2011} use the 2MRS galaxy catalogue for their galaxy overdensity sample and fit the inverse Tully-Fisher relation to spiral galaxies from the Spiral Field I-band ++ survey \citep[SFI++;][]{Masters2006, Springob2007} for their comparative velocity sample. They find a value of $f\sigma_8 = 0.31 \pm 0.04$, linearised by \citet{Carrick2015}. \citet{Carrick2015} themselves find $f\sigma_8 = 0.401 \pm 0.024$, using different variations of the 2MRS and SFI++ catalogues used by \citet{Davis2011}.

\citet{Boruah2020} utilise Type Ia supernovae as distance indicators to produce a catalogue of 465 peculiar velocities, which in combination with distances from SFI++ and 2MTF are used to estimate the peculiar velocity field. Comparing this to the velocity field reconstructed from the density field measured by 2M++ gives an estimate of $f\sigma_8 = 0.400 \pm 0.017$. They also produce an estimate of the growth rate by fitting to their SNe sample only (termed A2 in their work), finding $f\sigma_8 = 0.385 \pm 0.027$, but we do not show this in Figure \ref{fig:6dfcomp}.

\citet{Said2020} combine 6dFGSv and Sloan Digital Sky Survey \citep[SDSS;][]{York2000} peculiar velocity data to produce a joint analysis in order to measure the growth rate. Using Fundamental Plane peculiar velocities for 15 894 galaxies cumulatively from the 6dFGSv sample used by the other authors cited above, and the SDSS velocity sample \citep{Howlett2022}, and using the same 2M++ sample used by \citet{Boruah2020} for their galaxy overdensity sample. They find $f\sigma_8 = 0.311 \pm 0.027$.

Although there is some evidence to suggest that these results do not account for the error attributable to approximations in the reconstruction of the velocity field, and so underestimate their total error \citep{Lilow2021}, our results are also broadly consistent with each individual analysis.
\subsection{6dFGS Fisher matrix forecasts}
Finally, we can compare the results we find in this work to potential uncertainties predicted by Fisher matrix forecasts from 6dFGS data and future survey cross-correlation analyses. 

\citet{Koda2014}, using a 6dFGSv-like model survey, forecast an fractional uncertainty of $25$ per cent in measurements of $f\sigma_8$ from velocity data only. When extending this analysis to include galaxy density data -- a 'Two-field' constraint analogous with our cross-correlation analysis -- they forecast a fractional uncertainty of $15$ per cent in 6dFGS measurements of $f\sigma_8$. 

\citet{Howlett2017} also model constraints from 6dFGS-like redshift and velocity samples. They forecast a fractional uncertainty of $25.1$ per cent in measurements of $f\sigma_8$ for a velocity auto-correlation analysis. When including the information in the galaxy overdensity field this uncertainty is reduced to $11.2$ per cent. 

In comparison, the uncertainty in our measurement of $f\sigma_8$ is approximately $19$ per cent. Although these forecast errors are representative of our measurement, there are differences in detail caused by our use of configuration-space versus Fourier-space statistics, the exact selection function of the surveys, and modelling of the nuisance parameters.

Both authors also predict the potential uncertainties on $f\sigma_8$ and $\beta$ from combining future surveys such as the proposed TAIPAN galaxy survey \citep{DeCunha2017} and the combined Widefield ASKAP L-band Legacy All-sky Blind surveY \citep[WALLABY;][]{Koribalski2020} and proposed Westerbork Northern Sky HI Survey (WNSHS). Assuming $k_{\rm max} = 0.1 \, h$ Mpc$^{-1}$, \citet{Koda2014} forecast that $f\sigma_8$ and $\beta$ can be measured with $3.5$ per cent and $3.8$ per cent uncertainty respectively from the combined WALLABY and WNSHS surveys, using linear power spectra. \citet{Howlett2017} extend this, combining the WALLABY and WNSHS datasets with the TAIPAN survey. They find that by combining these three future surveys that $f\sigma_8$ can potentially be measured with $2.8$ per cent uncertainty. Looking further ahead to surveys such as DESI or 4MOST, which will provide us with potentially tens of millions of redshifts and on the order of one million velocity measurements, we should see these constraints tightened even further. At such levels of accuracy, it will be possible to place extremely tight constraints on possible modifications to General Relativity, or alternate theories of Dark Energy. 

\section{Conclusions}
\label{sec:conclusion}

In this paper, we present a self-consistent correlation function analysis of the joint galaxy density and peculiar velocity datasets of the 6-degree Field Galaxy Survey, and corresponding mock catalogues.  

This study expands the analysis performed in \citet{Turner2021} by modelling the effects of redshift space distortions and incorporating them into our existing framework. We show how the ensemble of models utilised in \citet{Turner2021} can be modified via multipole expansion and rewritten in terms of the power spectrum multipoles in order to capture the effects of distortions in galaxy clustering along the line of sight in redshift space, producing models for the dipole and octupole of the galaxy auto-correlation function and for the monopole and quadrupole of the galaxy-velocity cross-correlation function. We also demonstrate how these new models can be used in tandem to further improve the statistical uncertainty in measurements of cosmological parameters, by applying this new ensemble to the 6dFGS dataset. 

Similarly, we modify the estimators used in \citet{Turner2021} in order to fully capture the cosmological information in redshift space. The logarithmic distance ratio parameter $\eta$ is used in place of the regular velocity parameter for its Gaussian-distributed uncertainties, manifesting in the addition of a scaling factor to our estimators, modifications to the normalisation constants applied to our pair counts and a revision to the optimal FKP weighting used for our velocity sample. An FKP-style weighting is also computed for our galaxy sample to match that used in the velocity sample.

This method is first tested and verified on 6dFGS mock catalogues to demonstrate the unbiased recovery of the fiducial cosmology used to construct the 6dFGS mocks. Applying this method to our five-statistic ensemble we find $f\sigma_8 = 0.391 \pm 0.079$ ($20.2$ per cent statistical accuracy) and $\beta = 0.394 \pm 0.087$ ($22.1$ per cent statistical accuracy), within one standard deviation of the fiducial cosmology we expect from the mock data. We then apply the framework to the 6dFGS survey, using the velocity sample (6dFGSv) for our peculiar velocities and the redshift sample (6dFGSz) for our galaxy overdensities. 

After demonstrating that the method we've constructed is robust against our choice of fitting range, and assuming a fiducial value of $\sigma_8$, we produce measurements of $f\sigma_8$ and $\beta$ from the one-dimensional marginalised posteriors of the three-dimensional joint fit to ($f$, $b$, $\sigma_v$). We find $f\sigma_8 = 0.358 \pm 0.075$ and $\beta = 0.298 \pm 0.051$, estimating cosmological parameters with $20.9$ per cent and $21.8$ per cent statistical uncertainty respectively. 

Combining the 6dFGSv and 6dFGSz samples in a cross-correlation analysis provides tighter constraints on measurements of the growth rate than we would otherwise obtain from using either the 6dFGS velocity or redshift samples individually. We quantify this by directly contrasting the results we obtain from a fit considering all five estimators, to results we get from separately considering the galaxy auto-correlation multipoles (redshift data only) and from considering the velocity auto-correlation estimator plus the cross-correlation dipole (predominantly velocity data). The full consideration of all five estimators, ($\psi_1, \psi_2, \xi_{gu}^1, \xi_{gg}^0, \xi_{gg}^2$), produces constraints on $f\sigma_8$ which are $30.6$ per cent tighter than those we obtain from the RSD information only, via a fit to $\xi_{gg}^0$ and $\xi_{gg}^2$, and $43.2$ per cent tighter than those we obtain from velocity information, via a fit to $\psi_1$, $\psi_2$ and $\xi_{gu}^1$. Similarly, the constraints on $\beta$ from the full five-statistic fit are $35.0$ per cent and $59.4$ per cent tighter than we obtain from either the RSD or velocity information separately.

We compare this result with contemporary values of the local normalised growth rate from the literature, spanning different datasets, methods and techniques, finding that we are broadly in agreement with all of them. This comparison encompasses several different datasets, involving multiple methodological differences and various different techniques for extracting cosmological information from either the velocity field, the density field, or a combination of the two. Our measurement accuracies are also comparable to error forecasts of 6dFGS two-field results, whilst acknowledging some differences in approach. 

The grid-based analysis was selected due to its convenience, given we are interested in only three parameters ($f$, $b$, $\sigma_v$), and the running time for the process grows linearly with the number of parameter combinations we search through. It is also not computationally expensive in comparison to some of the results we have discussed, specifically those that require a re-computation of the covariance matrix at several steps. The analysis in its current state can be applied to larger datasets without issue because of this linear complexity, although more considered approaches to our parameter search could improve performance further. With the forecasts of \citet{Koda2014} and \citet{Howlett2017} in mind, upcoming PV surveys are expected to improve growth rate constraints to 3-5 per cent accuracy, providing stringent tests of the current cosmological model.

\section*{Acknowledgements}
We are grateful to Jun Koda, Paul Carter and Florian Beutler for providing the 6dFGS COLA mocks we employed in this paper. The 6dF Galaxy Survey was made possible by contributions from many individuals towards the instrument, the survey and its science. We particularly thank Matthew Colless, Heath Jones, Will Saunders, Fred Watson, Quentin Parker, Mike Read, Lachlan Campbell, Chris Springob, Christina Magoulas, John Lucey, Jeremy Mould, and Tom Jarrett, as well as the dedicated staff of the Australian Astronomical Observatory and other members of the 6dFGS team over the years.

We thank the anonymous referee for many helpful comments which improved the clarity and presentation of the work. RJT would like to acknowledge the financial support received through a Swinburne University Postgraduate Research Award throughout the creation of this work. RR acknowledges support from the Australian Government through the Australian Research Council’s Australian Research Council Laureate Fellowship funding scheme (project FL180100168). 

We have used R \citep{Rteam2021} for our data analysis, and acknowledge that the plots in this paper were generated with the use of the {\sc magicaxis} package \citep{magicaxis2019}.
\section*{Data Availability}

The data underlying this article will be shared on reasonable request to the corresponding author.



\bibliographystyle{mnras}
\bibliography{ref} 




\appendix
\section{Galaxy auto-correlation function multipoles}
\label{app:ggderiv}
In this Appendix we provide a complete derivation of the galaxy auto-correlation function multipoles in terms of the model power spectrum multipoles.  The auto-correlation function between galaxy overdensities $\delta_g$ at positions $\vec{x}$ and $\vec{x'}$ is 
\begin{equation}
    \langle \delta_g(\vec{x}) \delta_g(\vec{x'}) \rangle = \int \frac{d^3\vec{k}}{(2\pi)^3}\, P_{gg}(\vec{k})\,e^{i\vec{k}.\vec{r}} ,
\end{equation}
where $\vec{r} = \vec{x'} - \vec{x}$. The power spectrum P$_{gg}(\vec{k})$ can be written as a multipole expansion as a function of the angle to the local line of sight with direction $\hat{x}$,
\begin{equation}
    P_{gg}(\vec{k}) = \sum_{\ell} P^{\ell}_{gg}(k)L_{\ell}(\hat{k}.\hat{x}) ,
\end{equation}
where $P^{\ell}_{gg}(k)$ are the galaxy power spectrum multipoles given in Section \ref{ssec:rsdmod}. We also implement the plane-wave expansion,
\begin{equation}
    e^{i\vec{k}.\vec{r}} = \sum_{\ell}(2\ell + 1)\,i^{\ell}j_{\ell}(kr)L_{\ell}(\hat{k}.\hat{r}) .
\end{equation}
Substituting in these expressions, using $d^3\vec{k} = k^2\,dk\,\Omega_k$,
\begin{equation}
    \begin{aligned}
        \langle \delta_g(\vec{x}) \delta_g(\vec{x'}) \rangle = \frac{1}{(2\pi)^3} \int dk\,k^2&\,\sum_{\ell}(2\ell + 1)\,i^{\ell}j_{\ell}(kr)\sum_{\ell'}P^{\ell'}_{gg}(k)\\
        &\int d\Omega_k L_{\ell}(\hat{k}.\hat{r}) L_{\ell'}(\hat{k}.\hat{x}) .
    \end{aligned}
    \label{eq:pwex_auto}
\end{equation}
We then apply the spherical harmonic addition theorem,
\begin{equation}
    L_{\ell}(\hat{k}.\hat{r}) = \frac{4\pi}{2\ell + 1}\sum_m Y_{\ell m}(\hat{k})Y_{\ell m}^*(\hat{r}) ,
\end{equation}
to the two Legendre polynomials at the end of Equation \ref{eq:pwex_auto}, finding:
\begin{equation}
    \begin{aligned}
        \int& d\Omega_k L_{\ell}(\hat{k}.\hat{r})L_{\ell'}(\hat{k}.\hat{x}) = \\
        &\int d\Omega_k \frac{4\pi}{2\ell + 1} \sum_m Y_{\ell m}(\hat{k})Y_{\ell m}^*(\hat{r})
        \frac{4\pi}{2\ell' + 1}\sum_{m'}Y_{\ell' m'}^*(\hat{k})Y_{\ell' m'}(\hat{x})\\
        &= \frac{4\pi}{2\ell + 1} \frac{4\pi}{2\ell' + 1} \sum_m \sum_{m'} Y_{\ell m}^*(\hat{r})Y_{\ell' m'}(\hat{x})
        \int d\Omega_k Y_{\ell m}(\hat{k})Y_{\ell' m'}^*(\hat{k}) .
    \end{aligned}
    \label{eq:long_integral}
\end{equation}
The last integral in Equation \ref{eq:long_integral} is equal to $\delta_{\ell\ell'}\delta_{mm'}$, such that,
\begin{equation}
    \begin{aligned}
        \int d\Omega_k L_{\ell}(\hat{k}.\hat{x}) &= \delta_{\ell\ell'} (\frac{4\pi}{2\ell + 1})^2 \sum_m Y_{\ell m}^*(\hat{r})Y_{\ell m}(\hat{x})\\
        &= \frac{4\pi}{2\ell + 1}L_{\ell}(\hat{r}.\hat{x}) \delta_{\ell\ell'} ,
    \end{aligned}
\end{equation}
such that the expression for $\langle \delta_g(\vec{x}) \delta_g(\vec{x'}) \rangle$ becomes,
\begin{equation}
    \langle \delta_g(\vec{x}) \delta_g(\vec{x'}) \rangle = \frac{1}{2\pi^2} \int dk k^2 \sum_{\ell}i^{\ell}j_{\ell}(kr)P^{\ell}_{gg}(k)L_{\ell}(\hat{r}.\hat{x}).
    \label{eq:deltasub} 
\end{equation}
When we compare Equation \ref{eq:deltasub} with the definition of the multipole expansion,
\begin{equation}
    \langle \delta_g(\vec{x}) \delta_g(\vec{x'}) \rangle = \sum_{\ell} \xi^{\ell}_{gg}(r)L_{\ell}(\hat{r}.\hat{x}) ,
\end{equation}
we can identify the galaxy auto-correlation function multipoles in terms of the power spectrum multipoles,
\begin{equation}
    \xi^{\ell}_{gg}(r) = \frac{i^{\ell}}{2\pi^2} \int dk k^2 j_{\ell}(kr)P^{\ell}_{gg}(k) ,
\end{equation}
and thus we will have non-zero galaxy auto-correlation function multipoles for $\xi^{0}_{gg}(r)$, $\xi^{2}_{gg}(r)$ and $\xi^{4}_{gg}(r)$.

\section{Galaxy-velocity cross-correlation function multipoles}
\label{app:guderiv}
In this Appendix we provide a complete derivation of the galaxy-velocity cross-correlation function multipoles in terms of the model power spectrum multipoles. The cross-correlation function between galaxy overdensity $\delta_g$ at position $\vec{x}$ and radial peculiar velocity $u$ at position $\vec{x'}$ has the form,
\begin{equation}
    \langle \delta_g(\vec{x}) u(\vec{x'}) \rangle = iaH \int \frac{d^3\vec{k}}{(2\pi)^3}\frac{1}{k}P_{g\theta}(\vec{k})(\hat{k}.\hat{x}')e^{i\vec{k}.\vec{r}} ,
\end{equation}
where $\vec{r} = \vec{x'} - \vec{x}$. We apply the `local flat-sky approximation', which is that for any pair of points $\hat{k}.\hat{x}' \approx \hat{k}.\hat{x}$. We write the cross-power spectrum $P_{g\theta}(\vec{k})$ as a multipole expansion as a function of the angle to the local line of sight with direction $\hat{x}$,
\begin{equation}
    P_{g\theta}(\vec{k}) = \sum_{\ell}P^{\ell}_{g\theta}(k)L_{\ell}(\hat{k}.\hat{x}) ,
\end{equation}
where $P^{\ell}_{g\theta}(k)$ are the galaxy-velocity cross-power spectrum multipoles given in Section \ref{ssec:rsdmod}.
Substituting in this expression and the plane wave expansion, we find,
\begin{equation}
    \begin{aligned}
        \langle \delta_g(\vec{x}) u(\vec{x'}) \rangle = \frac{iaH}{(2\pi)^3} & \int dk k \sum_{\ell}(2\ell + 1) i^{\ell} j_{\ell}(kr)\sum_{\ell'} P_{g\theta}^{\ell'}(k)\\
        &\int d\Omega_k L_{\ell}(\hat{k}.\hat{r})L_{\ell'}(\hat{k}.\hat{x})(\hat{k}.\hat{x}) .
    \end{aligned}
\end{equation}
We can rewrite $L_{\ell'}(\hat{k}.\hat{x})(\hat{k}.\hat{x})$ as its own expansion over Legendre polynomials, using the formula $L_{\ell}(x)L_{\ell'}(x) = \sum^{\ell + \ell'}_{\ell' = |\ell - \ell'|}A^{\ell''}_{\ell, \ell'}L_{\ell''}(x)$, where 
\begin{equation}
    A^{\ell''}_{\ell  \ell'} = 
    \begin{pmatrix}
        \ell & \ell' & \ell''\\
        0 & 0 & 0
    \end{pmatrix}
    (2\ell'' + 1) ,
\end{equation}
and the matrix is a Wigner 3j-symbol. This gives us,
\begin{equation}
    L_{\ell'}(\hat{k}.\hat{x})(\hat{k}.\hat{x}) = L_{\ell'}(\hat{k}.\hat{x})L_1(\hat{k}.\hat{x}) = \sum^{\ell' + 1}_{\ell''=|\ell' - 1|}A^{\ell''}_{\ell', 1}L_{\ell''}(\hat{k}.\hat{x}) .
\end{equation}
We can then write the cross-correlation function in the form
\begin{equation}
    \begin{aligned}
        \langle \delta_g(\vec{x}) u(\vec{x'}) \rangle &= \frac{iaH}{2\pi^2}\int dk k \sum_{\ell}(2\ell + 1) i^{\ell}j_{\ell}(kr)\sum_{\ell'}P_{g\theta}^{\ell'}(k)\\
        &\sum^{\ell' + 1}_{\ell''=|\ell' - 1|}A^{\ell''}_{\ell', 1} \int d\Omega_k L_{\ell}(\hat{k}.\hat{r})L_{\ell''}(\hat{k}.\hat{x}) .
    \end{aligned}
\end{equation}
Following the same spherical harmonic addition theorem steps as in Appendix \ref{app:ggderiv}, we find,
\begin{equation}
    \begin{aligned}
        \langle \delta_g(\vec{x}) u(\vec{x'}) \rangle &= \frac{iaH}{2\pi^2} \int dk k \sum_{\ell} i^{\ell}j_{\ell}(kr)\sum_{\ell'}P_{g\theta}^{\ell'}(k)\\
        &\sum^{\ell' + 1}_{\ell''=|\ell' - 1|}A^{\ell''}_{\ell', 1} \delta_{\ell, \ell''} L_{\ell}(\hat{r}.\hat{x}) .
    \end{aligned}
\end{equation}
By similarly comparing with the definition of the multipole expansion we can identify the galaxy-velocity cross-correlation function multipoles in terms of the power spectrum multipoles,
\begin{equation}
    \xi^{\ell}_{gu}(r) = \frac{i^{\ell + 1}aH}{2\pi^2} \int dk k j_{\ell}(kr)\sum_{\ell'}P^{\ell'}_{g\theta}(k)\sum^{\ell' + 1}_{\ell''=|\ell' - 1|}A^{\ell''}_{\ell', 1} \delta_{\ell, \ell''} ,
\end{equation}
and this leads to non-zero galaxy-velocity cross-correlation multipoles when $\ell$ is odd,
\begin{equation}
    \xi^1_{gu}(r) = -\frac{aH}{2\pi^2}\int dk k j_1(kr)\left(P^0_{g\theta}(k) + \frac{2}{5}P^2_{g\theta}(k)\right) ,
\end{equation}
and,
\begin{equation}
    \xi^3_{gu}(r) = \frac{aH}{2\pi^2}\int dk k j_3(kr)\left(\frac{3}{5}P^2_{g\theta}(k) + \frac{4}{9}P^4_{g\theta}(k)\right) .
\end{equation}


\bsp	
\label{lastpage}
\end{document}